\documentclass[11pt,a4paper]{article}
\usepackage{jheppub}

\usepackage{amsmath}
\usepackage{epsfig,epsf}
\usepackage{graphicx}
\usepackage{verbatim}

\title{QCD Reggeon Calculus From KLWMIJ/JIMWLK Evolution: Vertices, Reggeization and All.}
\author[a]{Tolga Altinoluk,}
\author[b] {Carlos Contreras,}
\author[c]{Alex Kovner,}
\author[b,d]{Eugene Levin,}
\author[e]{\\ Michael Lublinsky} 
\author[e]{and Arthur Shulkin}

\affiliation[a]{
Departamento de F\'\i sica de Part\'\i culas and IGFAE, Universidade de Santiago de Compostela, E-15782 Santiago de Compostela, Galicia-Spain}
\affiliation[b] {Departamento de F\'\i sica, Universidad T\'ecnica
Federico Santa Mar\'\i a,
and
Centro Cient\'\i fico-Tecnol$\acute{o}$gico de Valpara\'\i so, Avda. Espa\~na 1680,
Casilla 110-V,  Valpara\'\i so, Chile}
\affiliation[c]{Physics Department, University of Connecticut, 2152 Hillside road, Storrs, CT 06269, USA}
\affiliation[d]{Department of Particle Physics,  Tel Aviv University , Tel Aviv 69978, Israel}
\affiliation[e]{Physics Department, Ben-Gurion University of the Negev, Beer Sheva 84105, Israel}
 
\date{\today}
\abstract{
We show explicitly how the high energy QCD evolution generated by the KLWMIJ Hamiltonian can be cast in the form of the QCD Reggeon Field Theory.
We show how to reduce the KLWMIJ Hamitonian to physical color singlet degrees of freedom. We suggest a natural way of defining the Pomeron and other Reggeons in the framework of the KLWMIJ evolution and derive the QCD Reggeon Field Theory Hamiltonian which includes several lowest Reggeon operators. This Hamiltonian generates  evolution equations for all Reggeons in the case of dilute-dense scattering, including the nonlinear  Balitsky-Kovchegov equation for the Pomeron. We also find explicit expressions for the Reggeon conjugate operators in terms of QCD operators, and derive their evolution equations. This provides a natural and unambiguous framework for reggeization procedure introduced in \cite{BW, BE}. The Bartels triple Pomeron vertex is inherited directly from the RFT Hamiltonian. For simplicity in the bulk of the paper we work in the large $N_c$ limit.
}

\date{\today}
\begin{document}
\maketitle

\pagestyle{empty}
\newpage

\mbox{}

\pagestyle{plain}

\setcounter{page}{1}
\section{Introduction}

Attempts to formulate QCD Reggeon calculus have a long history. First steps towards this goal were made in the first papers on low x physics \cite{GLR,MUPA}, and especially in the papers by Bartels and collaborators \cite{BW, BLW, Lotter, BV, BE,BBV} who derived the triple Pomeron vertex. This work however did not provide a framework or a practical direction of how to derive the actual Reggeon field theory, but
only considered the evolution equations for certain correlation functions. 
Furthermore, the derivation of the triple Pomeron vertex suffers from certain arbitrariness, as the procedure of constructing the reggeized parts of the Green's function was not sharply defined in these works and is not clearly process independent.
Later papers by Braun \cite{braun} propose a field theory description of the Pomeron interactions, but its relation to QCD  has not been rigorously derived.

 Several years ago  two of us have shown \cite{reggeon,yinyang} that the  Hamiltonian which governs the evolution of QCD amplitudes with energy in the so-called Color Glass Condensate (CGC) framework \cite{balitsky},\cite{kovchegov},\cite{jimwlk} should indeed be understood as the explicit QCD realization of the Reggeon Field Theory. This was supplemented by the analysis of gluon reggeization in the same framework in \cite{reggegluon}. The discussion in \cite{reggeon},\cite{reggegluon} was in terms of the eigenfunctions of the Hamiltonian, rather than the Hamiltonian itself. This setup is somewhat different from the one usually employed within the Reggeon Field Theory, which possibly obscured the connection to a certain extent.   In the present work we make an additional step to establish the direct interpretation of the JIMWLK/KLWMIJ Hamiltonian as the Reggeon Field Theory Hamiltonian. We provide the precise procedure of rewriting this Hamiltonian directly in the language of Reggeon fields. Here we deal with the KLWMIJ approximation to the RFT Hamiltonian\cite{klwmij}, the approximation valid for scattering of a dilute projectile on a dense target. We note that due to exact duality between the JIMWLK and KLWMIJ forms\cite{duality}, the exact same procedure can be applied to the JIMWLK limit. Although mathematically equivalent, it resums a different set of nonlinear corrections and thus is not physically equivalent. In future we intend to extend the analysis to include the more general RFT Hamiltonian. Following a lot of interesting work in the recent years \cite{msw},\cite{pomloops}, such a Hamiltonian which encompasses both, the JIMWLK and the KLWMIJ limits was derived in \cite{aklp}.

For simplicity, in the bulk of this paper we work in the leading order in $1/N_c$ expansion, but we provide some comments explaining how to extend this approach beyond the large $N_c$ limit. We stress, that the large $N_c$ limit we consider does not reduce to the dipole model, and already in this limit the relevant degrees of freedom are not only the Pomeron, but also higher color singlet operators. The Reggeon Hamiltonian includes these degrees of freedom even in the large $N_c$ limit. 

Rewriting KLWMIJ Hamiltonian in terms of reggeon fields 
\footnote{In this paper  by Reggeon we mean a colorless exchange amplitude, and not a reggeized gluon exchange in the adjoint representation.} requires us first to choose the reggeon degrees of freedom. There is a certain freedom of choice in this procedure, akin to freedom of definition of convenient field variables in an effective field theory. The choice is optimized by requiring simple implementation of the symmetries of the theory on the set of basics fields.
Our definition of the Pomeron field $P$ and other Reggeon fields  $B$ etc. is also informed by the weak field expansion in which the KLWMIJ Hamiltonian reduces to the BKP hierarchy\cite{bkp} generalization of the BFKL equation\cite{bfkl}. The number of such Reggeon fields in principle is infinite. We will restrict ourselves to the set of several lowest fields, which includes the Pomeron $P$, the Odderon $O$ and two more Reggeons, $B$ and $C$. The same procedure can be  straightforwardly applied for any extended set of Reggeon degrees of freedom. 

We pay particular attention to the definition of canonically conjugate fields $P^\dagger$. $B^\dagger$ etc. and their construction in terms of operators in the KLWMIJ Hilbert space. The evolution equations for $P$, $B$ etc. follow directly from the KLWMIJ Hamiltonian. We derive evolution equations of $P^\dagger$, $B^\dagger$ and their matrix elements. We explain that these evolution equations are equivalent (modulo a slight modification) to the evolution equations for functions $D^n$ introduced  in \cite{BW, BE}. We show explicitly how reggeization parts constructed in \cite{BW, BE} arise naturally when one relates $D^4$ to the matrix elements of the conjugate Reggeon operators. We demonstrate that the irreducible part of $D^4$ is precisely that part of $D^4$ which is given in terms of the matrix elements of $P^{\dagger 2}$ and $B^\dagger$, while the reggeization parts are given by admixture of $P^\dagger$ in $D^4$. We then explain the general algorithm of extracting reggeization corrections for an arbitrary $D^n$. 

The procedure of deriving the reggeization corrections that we discuss is fixed completely once we choose the field basis for representation of the Reggeon Field Theory. In this sense it rids the {\it ad hoc} procedure of  \cite{BW, BE} of the inherent arbitrariness\footnote{We note that a discussion of the Bartels' vertex in the framework of the Balitsky hierarchy was given in \cite{chirilli}. Our approach to this question is quite different and  more general.}.

Some of the material of this paper is not new and has already appeared in \cite{remarks},\cite{cut},\cite{gluons}. We include it here for completeness in order to present a coherent picture of the approach.

\section{The High Energy Evolution.}
In the high energy approximation any observable in a scattering process is calculated according to the following template
\begin{equation}\label{o}
\langle {\cal O}\rangle=\langle\int d\rho\delta(\rho)W[\delta/\delta\rho]\,{\cal O}[\rho,\alpha]\rangle_\alpha
\end{equation}
Here $\cal O$ is the observable in question, $W$ defines the probability density for the distribution of projectile charge density $\rho$ while $\alpha$ is the color field of the target. In the KLWMIJ approach the evolution to higher energy is given by evolving the functional $W$. The target field averaging procedure is not affected by the evolution and thus does not play a role in our discussion.

A particular example is an $S$-matrix of a projectile dipole. It is given by
\begin{equation}
S=\langle\int d\rho \delta(\rho)\frac{1}{N_c}tr[R^\dagger(x)R(y)]e^{i\int_z\rho(z)\alpha(z)}\rangle
\label{dipoleamp}
\end{equation}
where 
\begin{equation}
R(x)=e^{T^a\delta/\delta\rho^a(x)}
\end{equation}
with 
and $T^a$ - a generator of $SU(N_c)$ in the fundamental representation. 

The $S$-matrix operator is given by the eikonal expression $\hat S=e^{i\int_z\rho(z)\alpha(z)}$.
The unitary matrix $R(x)$ represents the scattering amplitude of a quark at the transverse coordinate $x$. This interpretation stems from the fact that the action of $\delta/\delta\rho^a(x)$ on the eikonal exponential in eq.({\ref{dipoleamp}) turns every factor of $\delta/\delta\rho^a(x)$ into $i\alpha^a$, thus turning $R$ into the eikonal scattering amplitude of a fundamental color charge in the background field of the target.
 
In general if the projectile is not a single dipole, the functional $W$ depends in some nontrivial way on a variety of color neutral objects
\begin{equation}
\frac{1}{N_c}tr[R^\dagger(x)R(y)]\rightarrow W[d(x_1,x_2),Q(x_1,x_2,x_3,x_4),...]
\end{equation}
where we have introduced
\begin{equation}\label{dipqua}
d(x,y)\equiv \frac{1}{N_c}tr[R(x)R^\dagger(y)]; \ \ \ \ Q(x,y,u,v)\equiv \frac{1}{N}tr[R(x)R^\dagger(y)R(u)R^\dagger(v)];\ \ \ ...
\end{equation}
In the following we will refer to $d$ as a dipole and $Q$ as a quadrupole. The physical meaning of $d$ is that of a scattering matrix of a single dipole, while $Q$ is the interference term in the scattering of a two dipole state (we will however refer to it as a quadrupole, following the time honored tradition). We note that although at large $N_c$ the contribution of $Q$ to the scattering amplitude of a two dipole state is subleading, it does contribute at leading order to other observables, as for example, to the inclusive two- and higher gluon production\cite{cut},\cite{gluons}. Thus even in the large $N_c$ limit one has to consider $Q$ as well as other, more general color singlet amplitudes.

The eikonal factor $\exp \{i\rho\alpha\}$ can be expanded in powers of the field $\alpha$. Each power of $\alpha$ corresponds to an exchange of a $t$-channel gluon\cite{reggeon}. Perturbation theory in $\alpha$ breaks down when the target field is large, which is the situation where KLWMIJ evolution is applicable. Nevertheless, the perturbation theory is helpful in illustrating some features of our approach. In particular a given power in $\alpha$ multiplies the matrix element of a given power of the color charge density. Thus matrix elements of the type
\begin{equation}\label{matr}
D_W^n=\int d\rho\delta(\rho)W[\delta/\delta\rho]\rho^{a_1}(x_1)...\rho^{a_n}(x_n)
\end{equation}
are important objects to consider. As we will show below these are important elements in the QCD Reggeon field theory.

The evolution of these matrix elements is given by the action of the KLWMIJ Hamiltonian \cite{klwmij}
\begin{equation}
\frac{dD^n_W}{dY}=\int d\rho\delta(\rho) W[\delta/\delta\rho]H_{KLWMIJ}\rho^{a_1}(x_1)...\rho^{a_n}(x_n)
\label{mevol}
\end{equation}
Here
\begin{equation}
H_{KLWMIJ}=\frac{\alpha_s}{2\pi^2}\int_{x,y,z}{K_{xyz}\left\{J^a_L(x)J^a_L(y)+J^a_R(x)J^a_R(y)-2J^a_L(x)R^{ab}_zJ^b_R(y)\right\}}
\label{klwmij}
\end{equation}
with the kernel
\begin{equation}
K_{x,y;z}=\frac{(x-z)_i(y-z)_i}{(x-z)^2(y-z)^2}
\end{equation}
The left and right rotation generators when acting on functions of $R$ have the representation
\begin{eqnarray}\label{LR}
J^a_L(x)=tr\left[\frac{\delta}{\delta R^{T}_x}T^aR_x\right]-tr\left[\frac{\delta}{\delta R^{*}_x}R^\dagger_xT^a\right]  \\  
J^a_R(x)=tr\left[\frac{\delta}{\delta R^{T}_x}R_xT^a\right] -tr\left[\frac{\delta}{\delta R^{*}_x}T^aR^\dagger_x\right]
\end{eqnarray}
We also note, that when $H_{KLWMIJ}$ acts on gauge invariant operators (operators invariant under $SU_L(N)$ and $SU_R(N)$ rotations), the kernel $K_{xyz}$ can be substituted by the so called dipole kernel
\begin{equation}
K_{x,y;z}\rightarrow -\frac{1}{2}M_{x,y;z}; \ \ \ \ \ \ \ \ M_{xy;z}=\frac{(x-y)^2}{(x-z)^2(y-z)^2}
\end{equation}

As discussed at length in the literature, the Hamiltonian $H_{KLWMIJ}$ in eq.(\ref{mevol}) can be either understood as acting to the left on $W$, or to the right on the product of the charge densities. When acting on a function of $\rho(x)$, the left and right charge densities in eq.(\ref{klwmij}) can be conveniently represented as\footnote{Eq.(\ref{js}) is the generalization of formulae given in \cite{dipoles}. In \cite{dipoles} we have given the Taylor expansion of eq.(\ref{js}) up to terms $\tau^4$. We are unaware of the all order resummed expressions eq.(\ref{js}) in the literature.} 
\begin{equation}\label{js}
J^a_L(x)=\rho^b(x)\left[\frac{\tau(x)}{2}\coth\frac{\tau(x)}{2}-\frac{\tau(x)}{2}\right]^{ba};\ \ \ \ J^a_R(x)=\rho^b(x)\left[\frac{\tau(x)}{2}\coth\frac{\tau(x)}{2}+\frac{\tau(x)}{2}\right]^{ba}
\end{equation}
where
\begin{equation}
\tau(x)\equiv t^a\frac{\delta}{\delta\rho^a(x)}
\end{equation}
with $t^a_{bc}= if^{abc}$ - the generator of $SU(N_c)$ in the adjoint representation.
We will use this explicit representation in subsequent sections.
\section{Elements of Reggeon Calculus: the Pomeron and Other Reggeons.}
As is clear from the previous section, and as was explicitly pointed out in \cite{reggeon}, the KLWMIJ Hamiltonian defines a 2+1 dimensional quantum field theory. The Hilbert space of this theory is spanned by functionals $W[R]$. The complete set of operators is spanned by the unitary matrices $R(x)$ and the generators of either left or right local  $SU(N_c)$ transformation $J_L^a(x)$ or $J_R^a(x)$. The local $SU(N_c)$ generators play the role of the canonical conjugates to the unitary matrix $R(x)$.

This quantum field theoretical structure is very similar to the one we expect from the QCD Reggeon Field Theory. It is however not quite identical. The Reggeons are physical scattering amplitudes, and thus have to be gauge invariant. The KLWMIJ Hamiltonian in principle allows one to consider color nonsinglet exchanges, although as has been shown in many cases\cite{cut}, and as we expect in general, color nonsinglet amplitudes will vanish due to infrared divergencies. This is the same fate as befalls the reggeized gluon\cite{reggluon}. There remains therefore additional step to be made in order that we can identify KLWMIJ with the RFT. Namely we need to project the KLWMIJ Hamiltonian on the $SU_L(N_c)\times SU_R(N_c)$ invariant subspace. This is the right thing to do, since the left index on the matrix $R$ correspond to the color state of the initial state, while the right index to the color of the final state. Thus color singlet exchange contributions to scattering of color singlet states are spanned by the $SU_L(N_c)\times SU_R(N_c)$ observables.

The aim of this and the next section is to perform this projection and obtain the Hamiltonian which governs the evolution of the color singlet amplitudes only.

Such an undertaking is similar to deriving a low energy effective theory for QCD in terms of physical states - pions, nucleons etc.
 In fact the Reggeon calculus in a quirky way does correspond to the low energy limit of the high energy QCD. It is best suited to the regime where a small number of Reggeon exchanges dominate the scattering, which means that the value of the scatterin matrices $d$, $Q$ etc. are close to unity. In this regime the Hamiltonian can be expanded in powers of $1-d$, $1-Q$ etc. 
 
 Just like in derivation of the effective low energy theory for pions, the first step is to decide which "composite fields" to choose as a convenient basis. 
The choice of the basis fields is to a large degree arbitrary. However following two basic rules makes the form of the resulting dynamics simpler. First, one needs to include the fields that interpolate the lightest excitations, and second the fields should transform as irreducible representations of the symmetry group of the Hamiltonian. 

The analog of the lightest excitations are in our case the states that dominate the evolution at lower energies. By low energies we mean the regime where the present eikonal approach is already valid, but any nonlinear effects are still negligible, namely the regime where the KLWMIJ Hamiltonian can be approximated by the BFKL dynamics. The eigenstates in this regime are the solutions of the linear BKP equations with fixed number of $t$-channel gluons: the Pomeron, the Odderon and higher BKP states \cite{bkp}.

As for the symmetries, as discussed in detail in \cite{reggeon}, the KLWMIJ Hamiltonian possesses $SU_R(N)\times SU_L(N)$ continuous symmetry group $R\rightarrow URV$, as well as the discrete signature $Z_2$ symmetry, $R\rightarrow R^\dagger; \ \ J_L^a\rightarrow -J_R^a$ and the charge conjugation symmetry, $R\rightarrow R^*$.
Since we are interested in scattering of physical color neutral states, our basic fields should be singlets under $SU_R(N)\times SU_L(N)$, and be even or odd under the action of discrete symmetries. 

Thus our strategy is to construct singlet multilocal fields, which in the leading order in expansion in powers of $\delta/\delta\rho$ overlap with solutions of the BKP equation.
The simplest such pair of fields is the Pomeron and the Odderon
\begin{equation}\label{pomeron}
P(x,y)=\frac{1}{2N_c}\left({\rm tr} [2-R(x)R^\dagger(y)]-{\rm tr} [R(y)R^\dagger(x)]\right)\end{equation}
\begin{equation}\label{odderon}
O(x,y)=\frac{1}{2N_c}\left({\rm tr} [R(x)R^\dagger(y)]-{\rm tr} [R(y)R^\dagger(x)]\right)\end{equation}
The Pomeron is signature and charge conjugation even, while the Odderon is signature and charge conjugation odd. Note, that while the Pomeron starts with the term quadratic in $\delta/\delta\rho$, the Odderon starts with $(\delta/\delta\rho)^3$.

This exhausts the fields that can be constructed from two fields $R$.
The next color singlet field in the hierarchy is the quadrupole defined in eq.(\ref{dipqua}). By itself it is however not a convenient choice for a basic field, since it mixes with the Pomeron in the evolution. Another way of saying it is that the expansion of $Q$ in powers of $\delta/\delta\rho$ starts at order $(\delta/\delta\rho)^2$ rather than $(\delta/\delta\rho)^4$. This can be rectified by subtracting a linear in $P$ term. The resulting combination with required quantum numbers (signature and charge conjugation even) is 
\begin{eqnarray}\label{B-reggeon}
B(1,2,3,4)&=&\frac{1}{4}\left[4-Q(1,2,3,4)-Q(4,1,2,3)-Q(3,2,1,4)-Q(2,1,4,3)\right]\\
&-&\left[P_{12}+P_{14}+P_{23}+P_{34}-P_{13}-P_{24}\right]\nonumber
\end{eqnarray}

The field $B$ defined in this way does not mix with the single Pomeron operator. In order $(\delta/\delta\rho)^4$ it does contain terms of the type $\frac{\delta}{\delta\rho^a(x)}\frac{\delta}{\delta\rho^a(y)}\frac{\delta}{\delta\rho^b(u)}\frac{\delta}{\delta\rho^b(v)}$, which are the same as the leading order expansion of two Pomerons. One can further refine the definition of $B$ by subtracting these double Pomeron terms, so that the resulting field is orthogonal to the two Pomeron operator. We find however, that this does not lead to any advantages, since even after the subtraction, the energy evolution mixes the operator $B$  with $P^2$. 
We describe the pertinent subtraction in the Appendix, but in the rest of this paper we use the field $B$ defined in eq.(\ref{B-reggeon}).

There are three independent charge conjugation even B-reggeons associated with given four points:
\begin{equation}
B(1,2,3,4); \ \ B(1,3,2,4); \ \ B(1,3,4,2)
\end{equation}
 Additionally there are three independent signature even, charge conjugation odd operators $C(1,2,3,4); \ \ C(1,3,2,4); \ \ C(1,3,4,2)$ defined as:
 \begin{equation}
 C(1,2,3,4)=\frac{1}{4}\left[Q(1,2,3,4)+Q(4,1,2,3)-Q(3,2,1,4)-Q(2,1,4,3)\right]
 \end{equation}
 The operators $C$ do not require any subtractions, since they are charge conjugation odd, while any possible subtraction must involve only Pomerons, which are charge conjugation even.
 
There is a total of nine independent operators that in the lowest order start with the fourth power of $\delta/\delta\rho$: $P_{12}P_{34};\ \ $ $ P_{13}P_{24}; \ \ $  $P_{14}P_{23}; \ \ $ $ B(1,2,3,4); \ \ $ $B(1,3,2,4);\ \ $ $B(1,3,4,2); \ \  $
\newline
$C(1,2,3,4); \ \ C(1,3,2,4); \ \ C(1,3,4,2)$. This indeed is the correct number of independent color singlets that can be constructed from four $t$-channel gluons.

The operators $B$ and $C$ are signature even. At the next level one should also consider signature odd operators
\begin{equation}
D^\mp(1,2,3,4)=\frac{1}{4}\left[Q(1,2,3,4)-Q(4,1,2,3)\right]\mp \frac{1}{4}\left[Q(2,1,4,3)-Q(3,2,1,4)\right]
\end{equation}
Expansion of $D^\pm$ starts with $(\delta/\delta\rho)^5$.

 The operators $B,\ C,\ D^\pm$ is the complete set of operators based on the quadrupole $Q$.
In principle one has to continue this procedure and define an infinite tower of operators which interpolate BKP states with arbitrary number of gluons in the lowest order in $\delta/\delta\rho$. At the next level one should start with the operator containing six points
\begin{equation}\label{X}
X(1,2,3,4,5,6)=\frac{1}{N_c}{\rm tr}(R_1R^\dagger_2R_3R^\dagger_4R_5R^\dagger_6)
\end{equation}
and project out of it all contributions that contain two or four gluons. In terms of the previously defined operators this amounts to subtraction of the terms of the type $P$, $P^2$ and $B$. 
Note that there are no $SU_L(N_c)\times SU_R(N_c)$ invariant operators with odd number of points. 
All these multipoint operators are independent, and all have to be included if one wants to consider high enough energies.

In this paper however, we will not dwell on the higher point operators any further and limit ourselves  to the set of operators $P$, $O$, $B$ and $C$ only. 

\section{Elements of Reggeon Calculus: the Evolution and the Hamiltonian.}
Our next task is to derive the dynamics of the Reggeons and to "`re-express" the KLWMIJ Hamiltonian in terms of the Reggeon degrees of freedom.  The simplest way of doing this, is by considering the action of the KLWMIJ Hamiltonian on the operators in question. For the operators $P$, $O$ and $Q$ this has been done in several papers \cite{remarks}, \cite{gluons}, \cite{cut}, \cite{quadrupole}.

Let us first concentrate on the dipole evolution.
Acting on the dipole operator by the KLWMIJ Hamiltonian, to leading order in $1/N_c$ one obtains
\begin{equation}\label{dipoleeq}
\frac{d}{ dY}d(x,y)=\frac{\bar \alpha_s}{2\,\pi}\,\int_{z} M_{x,y;z}\,[d_{x,z}\,d_{z,y}\,-\,d_{x,y}]
\end{equation}
Taking the real and imaginary parts of this equation we obtain
\begin{eqnarray}\label{pomeq}
\frac{d}{ dY}P_{x,y}&=&\frac{\bar \alpha_s}{2\,\pi}\,\int_{z} M_{x,y;z}\,[P_{x,z}+P_{z,y}\,-\,P_{x,y}-P_{x,z}P_{z,y}-O_{x,z}O_{z,y}];\\
\frac{d}{ dY}O_{x,y}&=&\frac{\bar \alpha_s}{2\,\pi}\,\int_{z} M_{x,y;z}\,[O_{x,z}+O_{z,y}\,-\,O_{x,y}-O_{x,z}P_{z,y}-P_{x,z}O_{z,y}]\nonumber
\end{eqnarray}
Now, suppose instead of acting on $P$ and $O$ we act on an arbitrary function $W[P,O]$. In the leading $N_c$ approximation we have
\begin{equation}
H_{KLWMIJ}W[P,O]=-\frac{d}{dY}W[P,O]=-\int_{x,y}\frac{d}{ dY}P_{x,y}\frac{\delta}{\delta P_{x,y}}W+\frac{d}{ dY}O_{x,y}\frac{\delta}{\delta O_{x,y}}W
\end{equation}
The subleading in $N_c$ terms were discussed in \cite{remarks} and we will comment on how to include these terms in our approach later. For now however, we concentrate on the large $N_c$ limit.
At large $N_c$ it is clear that when acting on a function of $P$ and $O$, the KLWMIJ Hamiltonian is equivalent to the sum of the following two operators 
\cite{LL,remarks}
\begin{eqnarray}\label{p}
H_P &=&-\frac{\bar \alpha_s}{2\,\pi}\,\int_{x,y,z} M_{x,y;z}\,\Big\{[P_{x,z}+P_{z,y}\,-\,P_{x,y}-P_{x,z}P_{z,y}-O_{x,z}O_{z,y}]\,P^\dagger_{x,y}\Big\}\\
H_O&=&-\frac{\bar \alpha_s}{2\,\pi}\,\int_{x,y,z} M_{x,y;z}\,\Big\{[O_{x,z}+O_{z,y}\,-\,O_{x,y}-O_{x,z}P_{z,y}-P_{x,z}O_{z,y}]\,O^\dagger_{x,y}\Big\}
\end{eqnarray}
Here the operators $P^\dagger$ and $O^\dagger$ are canonical conjugates to $P$ and $O$ and are defined via their action on functions of $P$ and $O$ as
\begin{equation}
P^\dagger_{xy}\equiv \frac{\delta}{\delta P_{xy}}W[P,O]; \ \ \ \ \  \ \ \ \ \ \ \ \ \ \ \ \ O^\dagger_{xy}\equiv \frac{\delta}{\delta O_{xy}}W[P,O]
\end{equation}
In the next section we will construct these operators explicitly in terms of the basic variables of $R$ and $J_{L(R)}$.

The action of $H_{KLWMIJ}$ on the quadrupole operator is also well known. The equation of motion satisfied by $Q$ is \cite{cut,gluons}
\begin{eqnarray}\label{quadrupoleeq}
\frac{d}{dY}Q(x,y,u,v)&=&\frac{\bar \alpha_s}{2\,\pi}\,\int_{z} -\,[M_{x,y;z}\,+\,M_{u,v;z}\,-\,L_{x,u,v,y;z}]
\,\,Q_{x,y,u,v}\,\\
&-&\, L_{x,y,u,v;z}\,d_{x,v}\,d_{u,y}\,-\,L_{x,v,u,y;z}\,d_{x,y}\,d_{u,v}
+\,L_{x,v,u,v;z}\,Q_{x,y,u,z}\,d_{z,v}\nonumber\\
&+&\,L_{x,y,x,v;z}\,Q_{z,y,u,v}\,d_{x,z}
\,+\,L_{x,y,u,y;z}\,Q_{x,z,u,v}\,d_{z,y}\,+\,L_{u,y,u,v;z}\,Q_{x,y,z,v}\,d_{u,z}\nonumber
\end{eqnarray}
where
\begin{eqnarray}
L_{x,y,u,v;z}\,&=&\,\left[\frac{(x\,-\,z)_i}{ (x\,-\,z)^2}\,-\,\frac{(y\,-\,z)_i}{(y\,-\,z)^2}\right ]\,\,
\left[\frac{(u\,-\,z)_i}{(u\,-\,z)^2}\,-\,\frac{(v\,-\,z)_i}{(v\,-\,z)^2}\right ]\nonumber \\
&=&\,\,\frac{1}{ 2}\,\left[\,M_{y,u;z}\,+\,M_{x,v;z}\,-\,M_{y,v;z}-M_{x,u;z}\,\right]
\end{eqnarray}
This leads to the following pair of equations for the operators $C$ and $B$
\begin{eqnarray}\label{cevolution}
\frac{d}{dY}C_{1234}&=&\frac{\bar \alpha_s}{2\,\pi}\,\int_{z}\Bigg\{-\,[M_{1,2;z}\,+\,M_{3,4;z}\,-\,L_{1,3,4,2;z}]
\,\,C_{1234}\\
&+&L_{1,4,3,4;z}C_{123z}+L_{1,2,1,4;z}C_{z234}+L_{1,2,3,2;z}C_{1z34}+L_{3,2,3,4;z}C_{12z4}\nonumber\\
&-&L_{1,4,3,4;z}C_{123z}P_{z4}-L_{1,2,1,4;z}C_{z234}P_{1z}-L_{1,2,3,2;z}C_{1z34}P_{z2}-L_{3,2,3,4;z}C_{12z4}P_{3z}\nonumber\\
&-&L_{1,4,3,4;z}D^+_{123z}O_{z4}-L_{1,2,1,4;z}D^+_{z234}O_{1z}-L_{1,2,3,2;z}D^+_{1z34}O_{z2}-L_{3,2,3,4;z}D^+_{12z4}O_{3z}
\Bigg\}\nonumber
\end{eqnarray}
\begin{eqnarray}\label{bevolution}
\frac{d}{dY}B_{1234}&=&\frac{\bar \alpha_s}{2\,\pi}\,\int_{z}\Bigg\{-\,[M_{1,2;z}\,+\,M_{3,4;z}\,-\,L_{1,3,4,2;z}]
\,\,B_{1234}\\
&+&L_{1,4,3,4;z}B_{123z}+L_{1,2,1,4;z}B_{z234}+L_{1,2,3,2;z}B_{1z34}+L_{3,2,3,4;z}B_{12z4}\nonumber\\
&-&L_{1,2,3,4;z}\Big[P_{1z}P_{2z}+P_{3z}P_{4z}\Big]-L_{3,2,1,4;z}\Big[P_{1z}P_{4 z}+P_{2z}P_{3z}\Big]\nonumber\\
&+&\Big(L_{1,2,3,4;z}+L_{1,4,3,2;z}\Big)\Big[P_{1z}P_{3z}+P_{2z}P_{4z}\Big]\nonumber\\
&-&L_{1,2,1,4;z}P_{1z}\left[P_{23}+P_{34}-P_{24}\right]-L_{1,2,3,2;z}P_{2z}\left[P_{14}+P_{34}-P_{13}\right]\nonumber\\
&-&L_{3,2,3,4;z}P_{3z}\left[P_{12}+P_{14}-P_{24}\right]-L_{1,4,3,4;z}P_{4z}\left[P_{12}+P_{23}-P_{13}\right]\nonumber\\
&+&L_{1,2,3,4;z}\Big(P_{14}P_{23}+O_{14}O_{32}\Big)+L_{1,4,3,2,z}\Big(P_{12}P_{34}+O_{12}O_{34}\Big)\nonumber\\
&-&L_{1,4,3,4;z}B_{123z}P_{z4}-L_{1,2,1,4;z}B_{z234}P_{1z}-L_{1,2,3,2;z}B_{1z34}P_{z2}-L_{3,2,3,4;z}B_{12z4}P_{3z}\nonumber\\
&-&L_{1,4,3,4;z}D^-_{123z}O_{z4}-L_{1,2,1,4;z}D^-_{z234}O_{1z}-L_{1,2,3,2;z}D^-_{1z34}O_{z2}-L_{3,2,3,4;z}D^-_{12z4}O_{3z}\nonumber\Bigg\}
\end{eqnarray}
Thus including $B$ and $C$ as additional arguments of $W$, the Hamiltonian $H_{KLWMIJ}$ acquires extra terms.
Using the symmetry properties of the operator $B^\dagger_{xyuv}$ and the coefficients $L$ these terms can be rewritten as
\begin{eqnarray}
H_C&=&-\frac{\bar \alpha_s}{2\,\pi}\,\int_{x,y,u,v,z}\Bigg\{-\,[M_{x,y;z}\,+\,M_{u,v;z}\,-\,L_{x,u,v,y;z}]
\,\,C_{xyuv}+4L_{x,v,u,v;z}C_{xyuz}C^\dagger_{xyuv}\nonumber \\
&-&4L_{x,v,u,v;z}C_{xyuz}P_{zv}C^\dagger_{xyuv}-4L_{x,v,u,v;z}D^+_{xyuz}O_{zv}C^\dagger_{xyuv}\\
&&\nonumber \\
H_B&=&-\frac{\bar \alpha_s}{2\,\pi}\,\int_{xyuvz}\Bigg\{-\,[M_{x,y;z}\,+\,M_{u,v;z}\,-\,L_{x,u,v,y;z}]
\,\,B_{xyuv}B^\dagger_{xyuv}+4L_{x,v,u,v;z}B_{xyuz}B^\dagger_{xyuv}\nonumber\\
&-&2L_{x,y,u,v;z}\Big[P_{xv}P_{uy}+O_{xv}O_{uy}\Big]B^\dagger_{xyuv}\nonumber\\
&-&2P_{xz}P_{yz}\Big[2L_{x,y,u,v;z}B^\dagger_{xyuv}-\Big(L_{x,u,y,v;z}+L_{x,v,y,u;z}\Big)B^\dagger_{xuyv}\Big]\nonumber\\
&-&4P_{xz}P_{yu}\Big[2L_{x,y,x,v;z}B^\dagger_{xyuv}-L_{x,y,x,u;z}B^\dagger_{xyvu}\Big]\nonumber\\
&-&4B_{xyuz}P_{zv}L_{x,v,u,v;z}B^\dagger_{xyuv}-4D^-_{xyuz}O_{zv}L_{x,v,u,v;z}B^\dagger_{xyuv}\Bigg\}
\end{eqnarray}
So that finally
\begin{equation}\label{klwmijnew}
H_{RFT}=H_P+H_O+H_B+H_C
\end{equation}
This defines the Hamiltonian of the QCD Reggeon Field Theory when restricted to act on the space of functionals of $P,\ O,\ B$ and $C$. 

A comment is in order here. One has to do a little more work in order to define a self consistent truncation of the RFT Hamiltonian. It is obvious from eq.(\ref{p}) that restriction to functions of $P$ and $O$ is consistent, since the Hamiltonian eq.(\ref{p}) does not contain any additional operators. The Hamiltonian eq.(\ref{klwmijnew}) on the other hand explicitly involves the operators $D^\pm$. Thus strictly speaking the restriction to the subspace of $P,\ O,\ B$ and $C$ is not entirely self consistent. To make it such, one would need to extend the present discussion and throw the functionals of $D^\pm$ into the fray. Since the evolution equation for the quadrupole operator closes, and $D^\pm$ is defined as a linear combination of the quadrupoles, it is obvious that including $D^\pm$ will make the system of equations for $P,\ O,\ B, \ C$ and $D^\pm$ closed. Thus such a truncation is self consistent in the large $N_c$ limit.  Including the evolution of $D^\pm$  is straightforward but we will not do it explicitly in the 

Hamiltonian eq.(\ref{klwmijnew}) has the typical fan diagram structure. Quadratic terms in the Hamiltonian generate homogeneous terms in the equations of motion for Reggeons. In particular the linear part of eq.(\ref{pomeq}) is the BFKL equation for the Pomeron and the Odderon, while the linear part of eqs.(\ref{cevolution},\ref{bevolution}) is the BKP equation for the compound state of four reggeized gluons\cite{bkp}. To be more precise one obtains the BKP equation as the evolution equation for $\nabla^1_1\nabla^2_2\nabla^2_3\nabla^2_4 B_{1234}$, rather than for $B_{12134}$ itself. The homogeneous part of eq.(\ref{bevolution}), in distinction from the standard large $N_c$ BKP kernel \cite{bkp},\cite{korchemsky},\cite{lipatovvega} contains terms of the type $M_{13;z}B_{123z}$  (and corresponding terms in eq.(\ref{cevolution})). These terms apparently involve interactions between non nearest neighbour gluons and are not present in the large $N_c$ BKP equations.
 Each one of these terms does not depend on one of the coordinates of the original reggeon field, and upon Fourier transform is propotional to delta function of one of the four momentum variables. Thus they indeed drop out of the BKP equation for $\nabla^1_1\nabla^2_2\nabla^2_3\nabla^2_4 B_{1234}$. Their presence in eq.(\ref{bevolution}) is necessary to ensure that the condition $B_{1134}=B_{1224}=B_{1233}=B_{1231}=0$  holds at all rapidities. 

The interaction terms in the Hamiltonian $H_{RFT}$ have the form of vertices, which all are of the "`triple Reggeon"' type $R_1\rightarrow R_2R_3$
(\ref{vertexfig}). In particular in addition to the standard triple Pomeron vertex $P\rightarrow PP$ eq.(\ref{klwmijnew}) gives rise to $P\rightarrow OO$; $O\rightarrow PO$; $B\rightarrow PP$; $B\rightarrow OO$;  $B\rightarrow PB$; $B\rightarrow D^-O$;  $C\rightarrow CP$ and $C\rightarrow D^+O$. All the vertices that are allowed by symmetries indeed appear in the Hamiltonian.
All these vertices are of the fan type, contributing to splitting of a Reggeon into two, but not to merging of two Reggeons into one. In a certain sense this is natural, as it is consistent with the folklore that KLWMIJ Hamiltonian contains only splittings but not mergings. 
\begin{figure}[ht]\epsfxsize=14cm
\centerline{\epsfbox{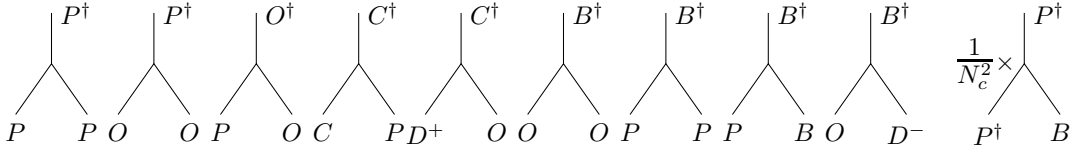}}
\caption{Triple Reggeon vertices derived from the KLWMIJ Hamiltonian}
\label{vertexfig}
\end{figure}

However in this context we want to note two things. Firstly, not all of these terms involve increase of the number of gluons in the t-channel. In particular the vertex $B\rightarrow PP$  contains a term, which in the leading order in $\delta/\delta\rho$ contributes to the homogeneous BKP equation, $(\delta/\delta\rho)^4\rightarrow (\delta/\delta\rho)^4$. Nevertheless when resummed into $B$ and $P$ it appears as part of the splitting vertex, rather than of the propagator of $B$.
Our second comment is, that the absence of the merging diagrams is only true in the leading order in $1/N_c$. As we will discuss in the last section, subleading in $1/N_c$ terms do indeed contain mergings of two Reggeons into one.

 We can now write down equations of motion for the conjugate operators. Those are generated by the action of the RFT Hamiltonian in the form eq.(\ref{klwmijnew}).
For example, acting on the conjugate Pomeron field we find
\begin{eqnarray}\label{conevol}
\frac{d}{dY}P^\dagger_{12}&=&\frac{\bar \alpha_s}{2\,\pi}\int_z \Big[M_{1,z;2}P^\dagger_{1z}+M_{2,z;1}P^\dagger_{2z}-M_{1,2;z}P^\dagger_{12}\\
&-&M_{1,z;2}P^\dagger_{1z}P_{2z}-M_{2,z;1}P^\dagger_{2z}P_{1z}-M_{2,z;1}O^\dagger_{z2}O_{z1}-M_{1,z;2}O^\dagger_{1z}O_{2z}\nonumber\\
&-&4L_{1uv2;z}B^\dagger_{1uv2}P_{uv}\nonumber\\
&-&4\Big(L_{1zuv;2}B^\dagger_{1zuv}P_{2z}+L_{2zuv;1}B^\dagger_{2zuv}P_{1z}\Big)+2\Big(L_{1uzv;2}+L_{1vzu;2}\Big)B^\dagger_{1uzv}P_{2z}\nonumber\\
&+&2\Big(L_{2uzv;1}+L_{2vzu;1}\Big)B^\dagger_{2uzv}P_{1z}-2\Big(L_{1z1v;2}B^\dagger_{1zuv}+L_{2z2v;1}B^\dagger_{2zuv}\Big)\Big(2P_{zu}-P_{zv}\Big)\nonumber\\
&+&4L_{u1u2;z}B^\dagger_{u1v2}P_{uz}-4\Big(L_{u1uv;z}B^\dagger_{u12v}+L_{u2uv;z}B^\dagger_{u21v}\Big)P_{uz}-2L_{z,2,u,2;1}C^\dagger_{zvu2}C_{zvu1}\nonumber\\
&-&2L_{z,1,v,1;2}C^\dagger_{zvu1}C_{zvu2}-2L_{z,2,u,2;1}B^\dagger_{zvu2}B_{zvu1}-2L_{z,1,v,1;2}B^\dagger_{zvu1}B_{zvu2}\Big]\nonumber
\end{eqnarray}
and similar equations for other conjugate operators. 

For the evolution of matrix elements of the type
\begin{equation}
\langle W|P^\dagger\rangle\equiv \int d\rho \delta(\rho)W[P,O...]P^\dagger
\end{equation}
only the first line in eq.(\ref{conevol}) is relevant. Whenever one has an operator $P$ or $O$ etc, which is built from derivatives $\delta/\delta\rho$ on the right of the last factor containing $\rho$, its matrix element vanishes, since it necessarily reduces to an integral of a derivative of a $\delta$ - function. All the factors of $\rho$ are contained in the conjugate Pomeron operator via the relation  eq.(\ref{solut}) and eq.(\ref{js}).  Thus for the purpose of calculating the evolution of this type of matrix elements one can use
\begin{equation}\label{conevol1}
\frac{d}{dY}P^\dagger_{xy}\approx \frac{\bar \alpha_s}{2\,\pi}\int_z \Big[M_{x,z;y}P^\dagger_{xz}+M_{y,z;x}P^\dagger_{yz}-M_{x,y;z}P^\dagger_{xy}\Big]
\end{equation}
On the other hand the rest of the terms in eq.(\ref{conevol}) do contribute to evolution of matrix elements  of the type
\begin{equation}\label{matrrx}
\langle W|P^\dagger P^\dagger\rangle\equiv \int d\rho \delta(\rho)W[P,O...]P^\dagger P^\dagger; \ \ \ \ \ {\rm or}  \ \ \ \ \langle W|P^\dagger O^\dagger\rangle\equiv \int d\rho \delta(\rho)W[P,O...]P^\dagger O^\dagger \ \ \ \ etc.
\end{equation}
when the leftmost factor $P^\dagger$ is evolved.
Calculating the evolution of the two Pomeron operator explicitly we find
\begin{eqnarray}\label{twopomev}
\frac{d}{dY}\left(P^\dagger_{12}P^\dagger_{34}\right)&=&\frac{\bar \alpha_s}{2\,\pi}\Bigg\{\int_z\Bigg[\Big[M_{1z;2}P^\dagger_{1z}P^\dagger_{34}+M_{2z;1}P^\dagger_{2z}P^\dagger_{34}
+M_{3z;4}P^\dagger_{12}P^\dagger_{z4}+M_{4z;3}P^\dagger_{12}P^\dagger_{z3}\nonumber\\
&-&\left(M_{12;z}+M_{34;z}\right)P^\dagger_{12}P^\dagger_{34}\Big]
-2\,\Big[L_{1,4,3,2;z}B^\dagger_{1432}+L_{1,3,4,2;z}B^\dagger_{1342}\Big]\Bigg]\nonumber\\
&-&\frac{1}{2}\Big[M_{14;3}P^\dagger_{14}\delta_{23}+M_{24;3}P^\dagger_{24}\delta_{13}+M_{13;4}P^\dagger_{13}\delta_{24}+M_{23;4}P^\dagger_{23}\delta_{14}\Big]\nonumber\\
&+&\int_{u,v}\Bigg[\Big[\Big(L_{1,3,u,v;2}+L_{1,v,u,3;2}\Big)B^\dagger_{13uv}-\Big(L_{1,u,3,v;2}+L_{1,v,3,u;2}\Big)B^\dagger_{1u3v}\Big]\delta_{24}\nonumber\\
&+&\Big[\Big(L_{2,3,u,v;1}+L_{2,v,u,3;1}\Big)B^\dagger_{23uv}-\Big(L_{2,u,3,v;1}+L_{2,v,3,u;1}\Big)B^\dagger_{2u3v}\Big]\delta_{14}\nonumber\\
&+&\Big[\Big(L_{1,4,u,v;2}+L_{1,v,u,4;2}\Big)B^\dagger_{14uv}-\Big(L_{1,u,4,v;2}+L_{1,v,4,u;2}\Big)B^\dagger_{1u4v}\Big]\delta_{23}\nonumber\\
&+&\Big[\Big(L_{2,4,u,v;1}+L_{2,v,u,4;1}\Big)B^\dagger_{24uv}-\Big(L_{2,u,4,v;1}+L_{2,v,4,u;1}\Big)B^\dagger_{2u4v}\Big]\delta_{13}\Bigg]\nonumber\\
&-&\int_v\Bigg[\Big[2L_{1,3,1,v;2}B^\dagger_{134v}+2L_{1,4,1,v;2}B^\dagger_{143v}-\left[L_{1,3,1,4;2}+L_{1,4,1,3;2}\right]B^\dagger_{13v4}\Big]\nonumber\\
&+&\Big[2L_{2,3,2,v;1}B^\dagger_{234v}+2L_{2,4,2,v;1}B^\dagger_{243v}-\left[L_{2,3,2,4;1}+L_{2,4,2,3;1}\right]B^\dagger_{23v4}\Big]\nonumber\\
&+&\Big[2L_{3,1,3,v;4}B^\dagger_{312v}+2L_{3,2,3,v;4}B^\dagger_{321v}-\left[L_{3,1,3,2;4}+L_{3,2,3,1;4}\right]B^\dagger_{31v2}\Big]\nonumber\\
&+&\Big[2L_{4,1,4,v;3}B^\dagger_{412v}+2L_{4,2,4,v;3}B^\dagger_{421v}-\left[L_{4,1,4,2;3}+L_{4,2,4,1;3}\right]B^\dagger_{41v2}\Big]\Bigg]\Bigg\}+...\nonumber\\
\end{eqnarray}
Similarly for the evolution of $B^\dagger$-reggeon we have
\begin{eqnarray}\label{B-reggeonev}
\frac{d}{dY}B^\dagger_{1234}&=&\frac{\bar \alpha_s}{2\,\pi}\Bigg\{\int_z\Big[-\left(M_{12;z}+M_{34;z}-L_{1342;z}\right)B^\dagger_{1234}\\
&+&L_{1z3z;4}B^\dagger_{123z}+L_{2z4z;3}B^\dagger_{214z}+L_{1z3z;2}B^\dagger_{341z}+L_{2z4z;1}B^\dagger_{234z}\Big]\Bigg\}+...\nonumber
\end{eqnarray}
where the ellipsis denotes terms containing  factors of $P$, $O$ etc., which do not contribute to matrix elements containing a single operator $B^\dagger$ or $P^\dagger P^\dagger$. One can of course calculate explicitly the remaining part of the evolution of $P^\dagger P^\dagger$ and $B^\dagger$ directly applying the RFT Hamiltonian to these operators. Here we will not be interested in higher matrix elements, and thus the terms kept in eqs.(\ref{twopomev},\ref{B-reggeonev}) are sufficient for our purposes.

We note, that while in terms of Reggeons, equations of motion generated by $H_{RFT}$ contain splitting vertices $1\rightarrow 2$, in terms of the conjugate reggeons the vertices look like merging vertices $2\rightarrow 1$. Eq.(\ref{twopomev}) contains two such vertices, $P^\dagger P^\dagger\rightarrow P^\dagger$ and $P^\dagger P^\dagger\rightarrow B^\dagger$.

\section{The Conjugate Reggeons.}
The Hamiltonian derived in the previous section in principle achieves the goal we have posed for ourselves. It expresses the content of KLWMIJ evolution entirely in terms of color singlet operators and their conjugates. Thus any initial probability distribution $W[P,O,B,C]$ can be evolved in the leading $N_c$ approximation. 

One more point to clarify, is the definition of the Reggeon conjugate operators $P^\dagger,\ etc.$. So far we have defined them formally as functional derivatives acting on $W$. However it is instructive to express them explicitly in terms of the operators acting on the full Hilbert space. This will also give us tools to show the explicit relation between our current approach and that of  \cite{BW, BE}.

 It is quite clear that the conjugate operators are closely related to the generators of the right and left color rotations $J_L(N_c), \ J_R(N_c)$. In this section we provide explicit expressions for  $P^\dagger, \ O^\dagger,\ B^\dagger$ and $C^\dagger$ in terms of the color charge density operators $J_{L(R)}$. 
 
We start with the pair $P^\dagger$, $O^\dagger$ disregarding the others for the moment. 
Specifically we need to find fields that satisfy the following relations
\begin{equation}\label{con}
P_{xy}P^\dagger_{uv}=\delta^+[(uv)-(xy)];\ \ \ O_{xy}P^\dagger_{uv}=0; \ \ \ P_{xy}O^\dagger_{uv}=0; \ \ \ \ O_{xy}O^\dagger_{uv}=\delta^-[(uv)-(xy)]
\end{equation}
where 
\begin{equation}
\delta^\pm[(uv)-(xy)]=\frac{1}{2}[\delta^2(x-u)\delta^2(y-v)\pm\delta^2(x-v)\delta^2(y-u)]
\end{equation}
Note that we do not require the operator commutation relation. Instead in eq.(\ref{con}) the operators $P^\dagger$ and $O^\dagger$ are understood to act on $P$ and $O$ which have been constructed explicitly as functions of the unitary matrix $R$. This is equivalent to requirement that $P^\dagger$ and $O^\dagger$ have the following matrix elements
\begin{equation}
\int d\rho \delta[\rho]P_{xy}P^\dagger_{uv}\equiv\langle P_{xy}|P^\dagger_{uv}\rangle=\delta^+[(uv)-(xy)]; \ \ \int d\rho \delta[\rho]O_{xy}P^\dagger_{uv}\equiv\langle O_{xy}|P^\dagger_{uv}\rangle=0; \ \ \ etc.
\end{equation}
Our reason for imposing only this "`weak"' commutation relation is that the conjugate operators only  appear in the matrix elements of the type eq.(\ref{o}). Thus only the matrix elements of the type 
$\langle W[P,O]|P^\dagger\rangle$ etc. are ever relevant and "`strong"' operator commutation relation is not required.

The above relations must hold for all $u\ne v$ and $x\ne y$. The definition of $P_{xy}$ and $O_{xy}$ is such that $P_{xx}=O_{xx}=0$. Thus the conjugate momenta should also be defined such that $P^\dagger_{xx}=O^\dagger_{xx}=0$.

Our strategy for establishing the operator form of the conjugates will be the following. We will consider the action of simple operators with correct quantum numbers on $P$ and $O$ and then solve for $P^\dagger$ and $O^\dagger$ expanding in powers of $P$ and $O$.
The simplest such relation is
\begin{eqnarray}\label{jj}
J_L^a(1)J_L^a(2)d(1,2)&=& J_R^a(1)J_R^a(2)d(1,2)=-\Lambda^4 \frac{N_c}{2}d(1,2)\nonumber\\
 J_L^a(1)J_L^a(1)d(1,2)&=&J_R^a(1)J_R^a(1)d(1,2)=\Lambda^4\frac{N_c}{2}d(1,2)
\end{eqnarray}
The dimensionfull constant $\Lambda$ is the UV cutoff, reflecting the fact that $J$ is not a charge, but a charge density. Formally the ultraviolet cutoff $\Lambda$ is defined as 
$\Lambda^4=[\delta^2(0)]^2 $ if $\delta(x)$ is properly regularized in the UV. 

Eq.(\ref{jj}) can be written as 
\begin{equation}\label{2j}
\frac{1}{2N_c}\Bigg[J_L^a(1)J_L^a(2)+J_R^a(1)J_R^a(2)\Bigg]=P^\dagger_{12}\Big[1-P_{12}\Big]-O^\dagger_{12}O_{12}-\delta_{12}\int_x\Big[P^\dagger_{1x}\Big[ 1-P_{1x}\Big]-O^\dagger_{1x}O_{1x}\Big]
\end{equation}
The last term in this equation is necessary to correctly reproduce the action of $J_{L(R)}^a(1)J_{L(R)}^a(1)$, since by our definition $P^\dagger_{11}=0$.

To get another relation involving $P^\dagger$ and $O^\dagger$ we consider
\begin{eqnarray}
{\rm tr} [T^aT^bT^c]J_L^a(1)J_L^b(1)J_L^c(2) d(1,2)&=&{\rm tr} [T^aT^bT^c]J_R^a(1)J_R^b(1)J_R^c(2) d(1,2)=-\frac{N_c^2}{8}\Lambda^6d(1,2)\nonumber\\
{\rm tr} [T^aT^bT^c]J_L^a(1)J_L^b(1)J_L^c(2) d(2,1)&=&{\rm tr} [T^aT^bT^c]J_R^a(1)J_R^b(1)J_R^c(2) d(2,1)=\frac{N_c^2}{8}\Lambda^6d(2,1)\nonumber\\
\end{eqnarray}
This is equivalent to
\begin{eqnarray}\label{3j}
&&-\frac{2}{N_c^2}\Big\{
\Big[{\rm tr} [T^aT^bT^c]J_L^a(1)J_L^b(1)J_L^c(2)-{\rm tr} [T^aT^bT^c]J_L^a(2)J_L^b(2)J_L^c(1)\Big]+({\rm L\rightarrow R})\Big\}=\nonumber\\
&&= O^\dagger_{12}\Big[1-P_{12}\Big]-P^\dagger_{12}O_{12}
\end{eqnarray}
Equations(\ref{2j},\ref{3j}) can be inverted to give
\begin{eqnarray}\label{solut}
P^\dagger_{12}&=&\Big[J^2_{12}(1-P_{12})+J^3_{12}O_{12}\Big]\frac{1}{(1-P_{12})^2-O_{12}^2}+\frac{1}{2N_c}\delta_{12}\Big[J^a_L(1)\int^{'} _{x}\hspace{-.3cm}J^a_L(x)+J^a_R(1)\int^{'} _{x}\hspace{-.3cm}J^a_R(x)\Big]\nonumber\\
O^\dagger_{12}&=&\Big[J^3_{12}(1-P_{12})+J^2_{12}O_{12}\Big]\frac{1}{(1-P_{12})^2-O_{12}^2}
\end{eqnarray}
where we have defined
\begin{eqnarray}
J^2_{12}&\equiv& \frac{1}{2N_c}\Bigg[J_L^a(1)J_L^a(2)+J_R^a(1)J_R^a(2)\Bigg]\\
J^3_{12}&\equiv& -\frac{2}{N_c^2}\Big\{
\Big[{\rm tr} [T^aT^bT^c]J_L^a(1)J_L^b(1)J_L^c(2)-{\rm tr} [T^aT^bT^c]J_L^a(2)J_L^b(2)J_L^c(1)\Big]+({\rm L\rightarrow R})\Big\}\nonumber
\end{eqnarray}
The symbol $\int^{'}$ in eq.(\ref{solut}) means that an infinitesimal neighborhood around point $x=1$ is not included in the integration region, or, in other words,
$$
\int_x^{'}\,\delta_{x1}\,=\,0
$$
Note that in principle the operators $P$ and $O$ do not commute with $J^2_{12}$ and $J^3_{12}$ and thus the ordering in eq.(\ref{solut}) is important. 

In eq.(\ref{solut}) we have kept formally all orders of the reggeon fields $P$ and $O$. However, as noted earlier, the Reggeon Field Theory representation is most useful in the regime where the reggeon fields themselves are small. We know from the solution of BFKL equation, that in the linear regime the reggeon fields scale as powers of energy multiplied by powers of the strong coupling constant, e.g. $P\propto \alpha_s^2e^{4\ln 2 \alpha_sY}$. Thus for rapidities $Y<\frac{1}{\alpha_s}\ln \frac{1}{\alpha_s^2}$ the Pomeron (and other reggeons) are small and can serve as expansion parameters. In this regime our expressions can be expanded in powers of $P$, $O$ etc. In our analogy with low energy effective theory of QCD this expansion is akin to the chiral perturbation theory. We will refer to this expansion  as low energy expansion.
The expanded form of the conjugate operators will be useful in the next section.

To leading order in this expansion we have
\begin{eqnarray}
P^\dagger_{12}&=&\frac{1}{2N_c}\Bigg[J_L^a(1)J_L^a(2)+J_R^a(1)J_R^a(2)+\delta_{12}\Big[J^a_L(1)\int^{'} _{x}J^a_L(x)+J^a_R(1)\int^{'} _{x}J^a_R(x)\Big]\Bigg]\\
O^\dagger_{12}&=&-\frac{2}{N_c^2}\Big\{
\Big[{\rm tr} [T^aT^bT^c]J_L^a(1)J_L^b(1)J_L^c(2)-{\rm tr} [T^aT^bT^c]J_L^a(2)J_L^b(2)J_L^c(1)\Big]+({\rm L\rightarrow R})\Big\}+...\nonumber
\end{eqnarray}

We now follow a similar procedure to find conjugates to $B$ and $C$,
\begin{eqnarray}
B_{xyuv}\,B^\dagger_{1234}&=&\frac{1}{8} \left[ \delta_{x1}\delta_{y2}\delta_{u3}\delta_{v4}+\delta_{x2}\delta_{y3}\delta_{u4}\delta_{v1}+\delta_{x3}\delta_{y4}\delta_{u1}\delta_{v2}+\delta_{x4}\delta_{y1}\delta_{u2}\delta_{v3}\right.\nonumber \\
&&\left.+\delta_{x3}\delta_{y2}\delta_{u1}\delta_{v4}+\delta_{x2}\delta_{y1}\delta_{u4}\delta_{v3}
+\delta_{x1}\delta_{y4}\delta_{u3}\delta_{v2}+\delta_{x4}\delta_{y3}\delta_{u2}\delta_{v1}\right]\nonumber\\
C_{xyuv}\,C^\dagger_{1234}&=&\frac{1}{8} \left[ \delta_{x1}\delta_{y2}\delta_{u3}\delta_{v4}+\delta_{x2}\delta_{y3}\delta_{u4}\delta_{v1}+\delta_{x3}\delta_{y4}\delta_{u1}\delta_{v2}+\delta_{x4}\delta_{y1}\delta_{u2}\delta_{v3}\right.\nonumber \\
&&\left.-\delta_{x3}\delta_{y2}\delta_{u1}\delta_{v4}-\delta_{x2}\delta_{y1}\delta_{u4}\delta_{v3}
-\delta_{x1}\delta_{y4}\delta_{u3}\delta_{v2}-\delta_{x4}\delta_{y3}\delta_{u2}\delta_{v1}\right]
\end{eqnarray}
Just like for the Pomeron and Odderon, we define $B^\dagger_{1134}=B^\dagger_{1224}=B^\dagger_{1233}=B^\dagger_{1231}=0$, since the B-reggeon $B_{ijkl}$ vanishes if any two of the adjacent coordinates coincide.

The details of the calculation of $B^\dagger$ are given in the Appendix. The result  to leading accuracy in $1/N_c$ and to leading order in low energy expansion is
\begin{eqnarray}\label{j4b}
&&-\frac{1}{8}\Big[J^4_L(1,2,3,4)+J^4_L(2,1,4,3)+J^4_R(1,2,3,4)+J^4_R(2,1,4,3)\Big]=B^\dagger_{1234}\\
&&-\frac{1}{2}\delta_{12}\int_u\Big(B^\dagger_{1u34}+B^\dagger_{1u43}-B^\dagger_{14u3}\Big)-\frac{1}{2}\delta_{34}\int_u\Big(B^\dagger_{123u}+B^\dagger_{12u3}-B^\dagger_{132u}\Big)\nonumber\\
&&-\frac{1}{2}\delta_{23}\int_u\Big(B^\dagger_{13u4}+B^\dagger_{1u34}-B^\dagger_{134u}\Big)-\frac{1}{2}\int_u\delta_{14}\Big(B^\dagger_{123u}+B^\dagger_{132u}-B^\dagger_{12u3}\Big)\nonumber\\
&&+\delta_{12}\delta_{34}\int_{uv}B^\dagger_{1u3v}+\delta_{14}\delta_{23}\int_uB^\dagger_{1u2v}\nonumber\\
&&+\frac{1}{4}\delta_{14}\delta_{23}P^\dagger_{12}+\frac{1}{4}\delta_{1234}\int_uP^\dagger_{1u}-\frac{1}{8}\Big[\delta_{123}P^\dagger_{14}+\delta_{234}P^\dagger_{12}+\delta_{124}P^\dagger_{13}
+\delta_{134}P^\dagger_{23}\Big]\nonumber
\end{eqnarray}
where we have defined
\begin{equation}
J^4_L(1,2,3,4)\equiv \frac{8}{N_c^3}{\rm tr}(T^aT^bT^cT^d)J_L^a(1)J_L^b(2)J^c_L(3)J_L^d(4)
\end{equation}

The more complete expression without the simplification of low energy expansion can be found in the Appendix. A similar expression can be found for $C^\dagger$.

Strictly speaking introduction of $B$ and $C$ also alters subleading terms in eq.(\ref{solut}), since $J^2$ and $J^3$ do not annihilate $B$ and $C$. 
For example we have
\begin{eqnarray}
\frac{2}{N_c}\Big[J^a_L(1)J^a_L(2)+J^a_R(1)J^a_R(2)\Big]B_{1234}&=&-\Lambda^2\Bigg[B_{1234}+\Big(2P_{34}+P_{14}+P_{23}-P_{13}-P_{24}\Big)\nonumber\\
&&\hspace{3.4cm}- P_{12}P_{34}-O_{12}O_{34}\Bigg]\nonumber\\
\frac{2}{N_c}\Big[J^a_L(1)J^a_L(2)+J^a_R(1)J^a_R(2)\Big]C_{1234}&=&-\Lambda^2C_{1234}
\end{eqnarray}
And therefore eq.(\ref{2j}) is modified to
\begin{eqnarray}\label{21j}
\frac{1}{2N_c}\Bigg[J_L^a(1)J_L^a(2)+J_R^a(1)J_R^a(2)\Bigg]&=&P^\dagger_{12}\Big[1-P_{12}\Big]-O^\dagger_{12}O_{12}\nonumber\\
&-&\delta_{12}\int_x\Big[P^\dagger_{1x}\Big[ 1-P_{1x}\Big]-O^\dagger_{1x}O_{1x}\Big]\nonumber\\
&-2&\int_{uv}B^\dagger_{12uv}\Bigg[B_{12uv}+\Big(2P_{uv}+P_{1v}+P_{2u}-P_{1u}-P_{2v}\Big)\nonumber\\
&-&P_{12}P_{uv}-O_{12}O_{uv}\Bigg]-2\int_{uv}C^\dagger_{12uv}C_{12uv}
\end{eqnarray}
The extra terms are proportional to $B$ and $C$ and therefore are not important in the lowest order in the low energy expansion. 

To summarize, in leading order in the low energy expansion
\begin{eqnarray}\label{conjugates}
P^\dagger_{12}&=&\frac{1}{2N_c}\Big\{J_L^a(1)J_L^a(2)+J_R^a(1)J_R^a(2)+\delta_{12}\Big[J^a_L(1)\int^{'} _{x}J^a_L(x)+J^a_R(1)\int^{'} _{x}J^a_R(x)\Big]\Big\}\nonumber\\
O^\dagger_{12}&=&-\frac{2}{N_c^2}
{\rm tr} [T^aT^bT^c]\Big\{J_L^a(1)J_L^b(1)J_L^c(2)-J_L^a(2)J_L^b(2)J_L^c(1)+({\rm L\rightarrow R})\Big\}\nonumber\\
\\
B^\dagger_{1234}&=&-\frac{1}{N_c^3}{\rm tr}(T^aT^bT^cT^d)\Big\{J_L^a(1)J_L^b(2)J^c_L(3)J_L^d(4)+(1\leftrightarrow 2; \  3\leftrightarrow 4)+\ \ (L\rightarrow R)\Big\}\nonumber\\
C^\dagger_{1234}&=&-\frac{8}{N_c^3}{\rm tr}(T^aT^bT^cT^d)\Big\{J_L^a(1)J_L^b(2)J^c_L(3)J_L^d(4)-(1\leftrightarrow 2; \ 3\leftrightarrow 4)+\ \ (L\rightarrow R)\Big\}\nonumber\\
\end{eqnarray}

\section{Reggeization {\it a la} Bartels et.al.}
Bartels and collaborators \cite{BLW, Lotter, BV, BE} have derived the expression for the three Pomeron vertex by considering evolution of expectation value of product of color charge density operators in the wave function of a virtual photon. Our aim in this section is to show that all the results of \cite{BW, BE} are reproduced in the natural and simple way in our approach. In particular, the reggeized parts discussed in \cite{BW, BE} follow automatically once the conjugate operators $P^\dagger$, $B^\dagger$ etc. defined above are identified with the irreducible parts of the appropriate correlators of color charge densities. The extension of the results of \cite{BW, BE} to higher point function becomes an algorithmic procedure, and one can calculate the reggeized part of $D^6$, $D^8$ etc.  as well as the vertices in the evolution of these functions if one wishes to do so.

We start with the reminder of the approach and the results of \cite{BW, BE}.
Consider the expectation value of a string of color charge density operators in the wave function of a virtual photon (\ref{d2d4})
\begin{equation}\label{D}
(-i\,g)^n\langle \hat\rho^{a_1}(x_1)...\hat\rho^{a_n}(x_n)\rangle_\gamma\equiv D^n_{a_1...a_n}(x_1...x_n)
\end{equation}
\begin{figure}[ht]\epsfxsize=10cm
\centerline{\epsfbox{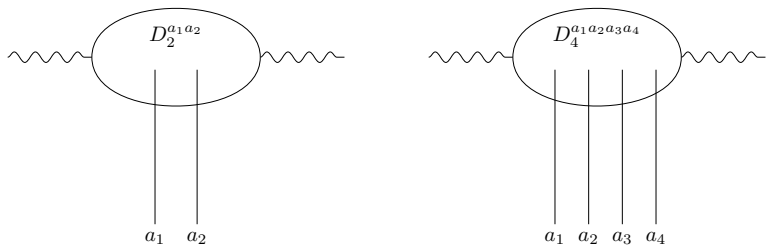}}
\label{d2d4}
\caption{$D^2$ and $D^4$}
\end{figure}
Note that the operators $\hat\rho^a(x)$ are not the same as the classical variable $\rho^a(x)$ that appears in functional integration in eqs.(\ref{o},\ref{matr}) and other equations in the previous sections of present paper. The operator $\hat\rho^a(x)$ is the quantum color charge density operator, and thus the ordering of operators in eq.(\ref{D}) is important whenever any two transverse coordinates coincide. This is an important point and we wil return to it later.

In the leading perturbative order the photon wave function is written as a charge conjugation even combination of color singlet dipoles with fundamental quark at transverse coordinate $x$ and antiquark at point $y$
\begin{equation}
|photon\rangle=\int_{x,y}F(x,y)\sum_{\alpha}\frac{1}{\sqrt N_c}|\alpha,x;\alpha,y\rangle
\end{equation}
with $F(x,y)=F(y,x)$. The form of the function $F(x,y)$ is not essential in the following, as it always factors out of all equations. The only important property of $|photon\rangle$ is that the state is charge conjugation even. For simplicity therefore we will consider just the charge conjugation even dipole state
\begin{equation}
\sum_{\alpha}\frac{1}{\sqrt 2N_c}\left[|\alpha,x;\alpha,y\rangle+|\alpha,y;\alpha,x\rangle\right]
\end{equation}

Ref.\cite{BW, BE} studied energy evolution of $D^n$ for $n=2,3,4,5$ and also considered $D^6$. The expressions of \cite{BW, BE} are all in momentum space, but we will transcribe them into coordinate space for convenience. The evolution of $D^2$ is simple. Writing $D^2_{ab}(12)\equiv \delta_{ab}D^2_{12}$ one has
\begin{equation}\label{d2ev}
\frac{d}{dY}D^2_{12}=\frac{\bar \alpha_s}{2\,\pi}\int_z \Big[M_{1,z;2}D^2_{1z}+M_{2,z;1}D^2_{2z}-M_{12;z}D^2_{12}-\delta_{12}\int_uM_{zu;1}D^2_{zu}\Big]
\end{equation}

 Further, \cite{bartels1}, \cite{BW} and \cite{BW, BE} were able to express the three point function at any rapidity in terms of the two point function
\begin{equation}\label{d3}
D^3_{a_1a_2a_3}(123)=\frac{g}{2}f^{a_1a_2a_3}\left[\delta_{12}D^2_{13}-\delta_{13}D^2_{12}+\delta_{23}D^2_{13}\right]
\end{equation}
To represent the evolution equation for the four point function in a simple way, \cite{BW, BE} split it into an "`irreducible part"' and a "`reggeized part"'
\begin{eqnarray}\label{reggeiz}
D^4_{a_1a_2a_3a_4}(1234)&=&D^{4I}_{a_1a_2a_3a_4}(1234)+D^{4R}_{a_1a_2a_3a_4}(1234); \\  &&\nonumber \\
D^{4R}_{a_1a_2a_3a_4}(1234)&=&-g^2\,d^{a_1a_2a_3a_4}\left[D^2_{14}\delta_{123}+D^2_{14}\delta_{234}-D^2_{12}\delta_{14}\delta_{23}\right]\nonumber\\
&&-g^2\,d^{a_2a_1a_3a_4}\left[D^2_{12}\delta_{134}+D^2_{13}\delta_{124}-D^2_{13}\delta_{12}\delta_{34}-D^2_{12}\delta_{13}\delta_{24}\right]\nonumber
\end{eqnarray}
with
\begin{equation}
d^{abcd}\equiv {\rm Tr}(T^aT^bT^cT^d)+{\rm Tr}(T^dT^cT^bT^a)
\end{equation}
The irreducible part $D^{4I}$ satisfies the equation
\begin{equation}
\frac{d}{dY}D^{4I}_{a_1a_2a_3a_4}(1234)=K\otimes D^{4I}\, +\, V\,\otimes\ D^2
\end{equation}
The first term in this equation schematically represents the action of the BFKL kernel on the four point function\footnote{Strictly speaking $KD^4$ contains not the BFKL kernel itself $K(x_1,x_2;y_1,y_2)$ but the BFKL kernel multiplied by the inverse of the two dimensional Laplacian for each outgoing coordinate $x_i$ and a factor of two dimensional Laplacian for each incoming index $y_i$. This kinematical factor is present since the original BFKL equation is satisfied by the color field $\alpha$ rather than color charge density $\rho$, while the relation between the two is $\alpha=\partial^2\rho$.}.
The reggeized part was defined in \cite{BW, BE} so that the irreducible four point function vanishes at initial rapidity. The subtraction of the reggeized part from $D^4$ is equivalent to the "`normal ordering"' of the operator $\hat\rho^4$ with respect to the dipole state.

The vertex operator $V$ in the momentum space has the form (here we follow the notations of Ref. \cite{BLV})
\begin{eqnarray}
V\,\otimes\ D^2&=&\,\frac{g^2}{2} 
\Bigg[\delta^{a_1a_2}\delta^{a_3a_4}
\Big[ G(k_1,k_2+k_3,k_4)+G(k_2,k_1+k_3,k_4)+G(k_1,k_2+k_4,k_3)+\nonumber\\
&&\hspace{2.3cm}+G(k_2,k_1+k_4,k_3)-G(k_1+k_2,k_3,k_4)-G(k_1+k_2,k_4,k_3)-\nonumber\\
&&\hspace{2.3cm}-G(k_2,k_1,k_3+k_4)-G(k_2,k_1,k_3+k_4)+
G(k_1+k_2,0,k_3+k_4)
\Big]\nonumber\\
&&\hspace{2.3cm}+(1\leftrightarrow 3)+(1\leftrightarrow 4)\Bigg]
\end{eqnarray}
The functions $G$ read
\begin{eqnarray}
G(k_1,k_2,k_3)&=& g^2\int \frac{d^2q_1d^2q_2}{(2\pi)^3} \,\delta^2(q_1+q_2-k_1-k_2-k_3)\,D^2(q_1,q_2)\,\nonumber \\
&\times&\Big[\frac{(k_2+k_3)^2}{(q_1-k_1)^2q_2^2}+
\frac{(k_2+k_1)^2}{(q_2-k_3)^2q_1^2}-
\frac{k_2^2}{(q_1-k_1)^2(q_2-k_3)^2}-
\frac{(k_1+k_2+k_3)^2}{q_1^2q_2^2}
\Big]\,-\nonumber \\ &&\nonumber \\
&-&(\omega(k_2)-\omega(k_2+k_3))D^2(k_1,k_2+k_3) 
-(\omega(k_2)-\omega(k_2+k_1))D^2(k_1+k_2,k_3)\nonumber\\
\end{eqnarray}
Here $\omega(k)=g^2/8\pi^3\ln k^2$ stands for the gluon trajectory. We now Fourier transform the above expressions. After noting the relation
\begin{eqnarray}
\int \frac{d^2k_1d^2k_2d^2k_3}{(2\pi)^6}\, e^{ik_1x_1+k_2x_2+k_3x_3}
G(k_1,k_2,k_3)&=&\frac{g^2}{8\pi^3}\,\big[M_{13,2}D^2_{13}-\int_xM_{1x,3}D^2_{1x}\delta_{23}-\nonumber\\
&-&\int_xM_{3x,1}D^2_{3x}\delta_{21}
+\int_{x,y}M_{xy,2}D^2_{xy}\delta_{123}\big]\nonumber\\
\end{eqnarray}
we arrive  at the expression for the vertex $V$ in coordinate space
\begin{eqnarray}\label{vertex}
V\,\otimes\ D^2&=&\frac{g^4}{16\,\pi^3} 
\Bigg[\delta^{a_1a_2}\delta^{a_3a_4}
\Big[M_{14;3}D^2_{14}\delta_{23}+M_{24;3}D^2_{24}\delta_{13}+M_{13;4}D^2_{13}\delta_{24}+M_{23;4}D^2_{23}\delta_{14}\nonumber \\
&-&M_{31;4}D^2_{31}\delta_{12}-M_{41;3}D^2_{41}\delta_{12}-M_{13;2}D^2_{13}\delta_{34}-M_{23;1}D^2_{23}\delta_{34}
\nonumber\\
&+&\int_x \Big[M_{3x;1}D^2_{3x}\delta_{124}+M_{4x;1}D^2_{4x}\delta_{123}+M_{1x;3}D^2_{1x}\delta_{234}+M_{2x;3}D^2_{24}\delta_{134}\nonumber \\
&+&M_{1x;3}D^2_{1x}\delta_{12}\delta_{34}+M_{13;x}D^2_{13}\delta_{12}\delta_{34}+M_{3x;1}D^2_{3x}\delta_{12}\delta_{34}+\int_y M_{xy;1}D^2_{xy}\delta_{1234}
\Big]\Big]\nonumber\\
&&+(1\leftrightarrow 3)+(1\leftrightarrow 4)\Bigg]
\end{eqnarray}
For compactness of notation we have introduced  $\delta_{123}\equiv \delta_{12}\delta_{23}$ etc.
Below we will establish a relation between the functions $D^n$ and the conjugate reggeons $P^\dagger$, $B^\dagger$. Once these relations are establish
the triple pomeron vertex operator $V$ could be read off directly from eq. (\ref{twopomev}). We find full agreement with the expression (\ref{vertex}).

Finally, the five point function behaves similarly to the three point function. Namely, at all rapidities it can be expressed as a given linear combination of the two- and four point functions.
\begin{eqnarray}\label{d5}
D^5_{a_1a_2a_3a_4a_5}(12345)&=&D^{5I}_{a_1a_2a_3a_4a_5}(12345)+D^{5R}_{a_1a_2a_3a_4a_5}(12345)\\ && \nonumber\\ 
D^{5R}_{a_1a_2a_3a_4a_5}(12345)&=&-g^3\,f^{a_2a_1a_3a_4a_5}\Big[D^2(12)\delta_{1345}-D^2_{15}\delta_{12}\delta_{345}\nonumber\\
&+&D^2_{13}\delta_{34}\delta_{125}-D^2_{12}\delta_{25}\delta_{134}\Big]\nonumber\\
&+&(2\leftrightarrow 4)+(2\leftrightarrow 3)\nonumber\\
&-&g^3\,f^{a_1a_2a_3a_4a_5}[D^2_{15}\delta_{1234}+D^2_{15}\delta_{2345}-D^2_{12}\delta_{15}\delta_{234}]\nonumber\\
D^{5I}_{a_1a_2a_3a_4a_5}(12345)&=&\frac{g}{2}\Big[f_{a_1a_2b}D^{4I}_{ba_3a_4a_5}(1345)\delta_{12}+ {\rm all \ permutations}\Big]\nonumber
\end{eqnarray}
where in the last line "`all permutations"' means all terms of similar type where the two coinciding coordinates take all possible values (the indices on $f$ are always in the ascending order while the index $b$ on $D^{4I}$ always takes place of the first missing $a_i$).
The five index tensor $f^{abcde}$ is defined as
\begin{equation}
f^{abcde}=\frac{1}{i}\Big[{\rm Tr}(T^aT^bT^cT^dT^e)-{\rm Tr}(T^eT^dT^cT^bT^a)\Big]
\end{equation}

The Reggeization according to Bartels et. al. therefore has two distinct elements. First, the odd point functions can be expressed entirely in terms of the even ones. Second, at least in the case of $D^4$ one has to perform "`normal ordering"' with respect to the dipole state. The normal ordered ("`irreducible"') piece of $D^4$ then satisfies an evolution equation with a simple $2\rightarrow 4$ vertex which has nice symmetry properties.

We will show in the following that both of these elements arise very simply in our approach. The first one is a direct consequence of the discrete signature symmetry of the KLWMIJ Hamiltonian, which guarantees the vanishing of an expectation value of any signature odd operator in any signature even state. Thus only the signature even parts of $D^3$, $D^5$ etc. do not vanish in the photon state. These signature even parts, as we explain below can be always expressed {\it operatorially} in terms of lower dimensional even point operators. This property is therefore immediately generalizable to all odd point functions beyond the ones considered in \cite{BW, BE}.

The normal ordering aspect of reggeization is synonymous with the statement that the most convenient set of variables are powers of $P^\dagger$, $B^\dagger$ etc. Since these operators are defined so that their matrix elements in a single dipole (Pomeron) vanish, they automatically correspond to normal ordered expressions when written in terms of the products of charge density operators $\rho^a$. Again this statement applies not only to $D^4$, but also gives a definite prescription of how to extend the procedure to $D^6$ etc, if so desired. We note that for $D^6$ this procedure leads to a combination of six-, four- and two point functions which vanish in one and two dipole states.

\subsection{Operator Symmetrization and the Odd-point Functions.}
First we have to establish the exact relation between the calculation of averages of the powers of the color charge density operator $\hat\rho^a(x)$ and matrix elements of the conjugate Reggeons. For this purpose it is most convenient to express both in terms of the classical variable $\rho^a(x)$ appearing in eqs.(\ref{o},{\ref{matr}) etc.
The distinct advantage of $\rho^a(x)$ as a basic variable, is that it has simple transformation properties under signature $\rho^a(x)\rightarrow -\rho^a(x)$.

 The relation between $\hat\rho^a(x)$ and $\rho^a(x)$ has been extensively discussed in \cite{dipoles}. The correlators of $\rho^a(x)$ are equal to the completely symmetrized correlators of $\hat\rho^a(x)$
\begin{equation}\label{hatr}
\langle \rho^{a_1}(x_1)...\rho^a_n(x_n)\rangle=\frac{1}{n!}\sum_{P(1,2...,n))}\langle \hat \rho^{a_{P_1}}(x_{P_1})...\hat\rho^{a_{P_n}}(x_{P_n})\rangle
\end{equation}
where summation goes over all permutations $P(1,...,n)$.
As shown in \cite{dipoles}, any product of $n$ operators $\hat\rho$ can be expressed as linear combinations of products of $\rho$ with $m\le n$ terms.
For example
\begin{equation}
\hat\rho^{a_1}_1\hat\rho^{a_2}_2=\rho^{a_1}_1\rho^{a_2}_2+t^{a_2}_{a_1c}\delta_{12}\rho^{c}_1
\end{equation}
\begin{eqnarray}
\hat\rho^{a_1}_1\hat\rho^{a_2}_2\hat\rho^{a_3}_3&=&\rho^{a_1}_1\rho^{a_2}_2\rho^{a_3}_3+\frac{1}{2}\bigg[t^{a_2}_{a_1c}\delta_{12}\rho^{c}_1\rho^{a_3}_3+t^{a_3}_{a_1c}\delta_{13}\rho^{c}_1\rho^{a_2}_2+t^{a_3}_{a_2c}\delta_{23}\rho^{a_1}_1\rho^{c}_2\bigg]\nonumber\\
&&+\frac{1}{12}\bigg[\{t^{a_2}t^{a_3}\}_{a_1c}-3(t^{a_3}t^{a_1})_{a_2c}\bigg]\delta_{12}\delta_{13}\rho^c_1
\end{eqnarray}
where $\{t^{a_2}t^{a_3}\}_{a_1c}=(t^{a_2}t^{a_3})_{a_1c}+(t^{a_3}t^{a_2})_{a_1c}$. These equations have to be understood under averaging sign like in eq.(\ref{hatr}), which we dropped for simplicity of notation.

The full expressions for symmetrization of the product of four and five operators $\hat\rho^{a_i}(x_i)$ is given in Appendix eqs.(\ref{appd4},\ref{appd5}) as they are too long to be reproduced in the body of the paper.
Thus the functions $D^n$ can be expressed in terms of the completely symmetrized correlators 
\begin{equation}
 \tilde D^n_{a_1...a_n}(x_1...x_n)\equiv\frac{(-i\,g)^n}{n!}\sum_{P(1,2...,n))}\langle \hat \rho^{a_{P_1}}(x_{P_1})...\hat\rho^{a_{P_n}}(x_{P_n})\rangle_\gamma
\end{equation}
For example for the three point function we have
\begin{equation}\label{ddtilde3}
D^3_{a_1a_2a_3}(123)=\tilde D^3_{a_1a_2a_3}(123)+\frac{g}{2}\bigg[f^{a_2}_{a_1c}\delta_{12}\tilde D^2_{ca_3}(13)+f^{a_3}_{a_1c}\delta_{13}\tilde D^2_{ca_2}(12)+f^{a_3}_{a_2c}\delta_{23}\tilde D^2_{a_1c}(12)\bigg]
\end{equation}
where we have neglected terms linear in $\rho$, since their average vanishes in any gauge invariant state.  

This relation is an operator relation, in the sense that it holds for averages in any gauge invariant state. On the other hand, the relations obtained in \cite{BW, BE} hold specifically for averages in (rapidity evolved) virtual photon state.
As mentioned earlier, the photon is a signature even state, while $\rho^3$ is a signature odd operator. Thus for a photon state  at any rapidity $\tilde D^3=0$. Additionally, in any gauge invariant state $\tilde D^2_{ab}=D^2\delta_{ab}$, and thus equation \ref{ddtilde3} for a photon directly reduces to eq.(\ref{d3}).

We have repeated this exercise for $D^5$ and have verified that eq.(\ref{d5}) arises in exactly the same way from the signature even terms in eq.({\ref{appd5})\footnote{ To do this exercise one has to be careful to properly identify the irreducible part  $D^{4I}$ in terms of $\tilde D^4$ and $D^2$.} The full expression for symmetrization of the five point function is given in the Appendix.
\subsection{The Irreducible Four Point Function and The Conjugate Reggeons.}
The four point function $D^4$ also contains a piece proportional to $D^2$ that arises due to symmetrization, eq. (\ref{appd4}). It is however not the case, that the complete reggeization part $D^{4R}$ as defined in \cite{BW, BE} arises due to symmetrization. The irreducible part $D^{4I}$ was defined in \cite{BW, BE} as a normal ordered with respect to the photon state. It therefore satisfies the initial condition $D^{4I}(Y_0)=0$. On the other hand there is no reason (symmetry or otherwise) for $\tilde D^4$ to vanish at initial rapidity, and indeed it does not. 

To determine the origin of the additional contributions to $D^{4R}$ we first rewrite the reggeization relation of \cite{BW, BE} in terms of the symmetrized four point function $\tilde D^4$. Using the formulae in the appendix and eq.(\ref{reggeiz}) we find the relation between the symmetrized four point function and the irreducible part of \cite{BW, BE}
\begin{eqnarray}\label{reggeiz1}
\tilde D^4_{a_1a_2a_3a_4}(1234)&=&D^{4I}_{a_1a_2a_3a_4}(1234)+\tilde D^{4R}_{a_1a_2a_3a_4}(1234); \\ \tilde D^{4R}_{a_1a_2a_3a_4}(1234)&=&-\frac{g^2}{3}T^{a_1a_2a_3a_4}\left[D^2_{14}\delta_{123}+D^2_{14}\delta_{234}+D^2_{12}\delta_{134}+D^2_{13}\delta_{124}\right]\nonumber\\
&+&\frac{g^2}{2}\Big[t^{a_1a_2a_3a_4}+t^{a_1a_3a_2a_4}+t^{a_1a_4a_3a_2}+t^{a_1a_4a_2a_3}\Big]D^2_{12}\delta_{14}\delta_{23}\nonumber\\
&+&\frac{g^2}{2}\Big[t^{a_1a_3a_4a_2}+t^{a_1a_2a_4a_3}+t^{a_1a_3a_2a_4}+t^{a_1a_4a_2a_3}\Big]D^2_{12}\delta_{13}\delta_{24}\nonumber\\
&+&\frac{g^2}{2}\Big[t^{a_1a_2a_3a_4}+t^{a_1a_3a_4a_2}+t^{a_1a_4a_3a_2}+t^{a_1a_2a_4a_3}\Big]
D^2_{13}\delta_{12}\delta_{34}\nonumber
\end{eqnarray}
with
\begin{equation}
t^{abcd}\equiv {\rm Tr} (T^aT^bT^cT^d)
\end{equation}
\begin{eqnarray}
T^{a_1a_2a_3a_4}&\equiv&{\rm Tr}(T^{a_1}T^{a_2}T^{a_3}T^{a_4})+ {\rm Tr}(T^{a_1}T^{a_2}T^{a_4}T^{a_3})+{\rm Tr}(T^{a_1}T^{a_3}T^{a_2}T^{a_4})\nonumber\\
&+&{\rm Tr}(T^{a_1}T^{a_3}T^{a_4}T^{a_2})+{\rm Tr}(T^{a_1}T^{a_4}T^{a_2}T^{a_3})+{\rm Tr}(T^{a_1}T^{a_4}T^{a_3}T^{a_2})
\end{eqnarray}
We stress that this relation between $\tilde D^4$ and $D^{4I}$ eq.(\ref{reggeiz1}) is completely equivalent to eq.(\ref{reggeiz}). We prefer to use it for the only reason, that is easy to obtain relation between $\tilde D^4$ and the conjugate Reggeons, as we will show below.

The object $\tilde D^4$ has four free indices. However we are only interested in its average in a generic gauge invariant state. Such an expectation value can be decomposed in terms of gauge singlet combinations. Defining (\ref{d4Pd4B})
\begin{equation}
\tilde D^4_P(1234)\equiv \tilde D^4_{aabb}(1234); \ \ \ \ \ \ \ \ \tilde D^4_B(1234)\equiv t^{a_1a_2a_3a_4}\tilde D^4_{a_1a_2a_3a_4}(1234)
\end{equation}
\begin{figure}[ht]\epsfxsize=10cm
\centerline{\epsfbox{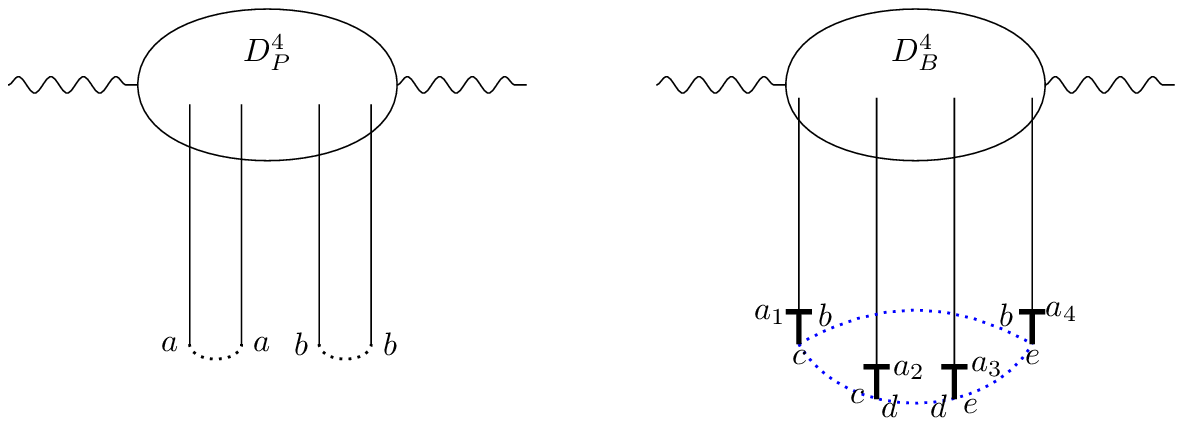}}
\label{d4Pd4B}
\caption{$D^4_P$ and $D^4_B$}
\end{figure}
to leading order in $1/N_c$ one has the following decomposition
\begin{eqnarray}\label{decomp}
&&\tilde D^4_{a_1a_2a_3a_4}(1234)=\nonumber\\
&=&\frac{1}{N^4}\Bigg[\delta^{a_1a_2}\delta^{a_3a_4}\Big[\tilde D^4_P(1234)-\frac{4}{N}[\tilde D^4_B(1234)+\tilde D^4_B(1432)+\tilde D^4_B(1243)+\tilde D^4_B(1342)]\Big]\nonumber\\
&+&\delta^{a_1a_4}\delta^{a_3a_2}\Big[\tilde D^4_P(1423)-\frac{4}{N}[\tilde D^4_B(1432)+\tilde D^4_B(1234)+\tilde D^4_B(1423)+\tilde D^4_B(1324)]\Big]\nonumber\\
&+&\delta^{a_1a_3}\delta^{a_2a_4}\Big[\tilde D^4_P(1324)-\frac{4}{N}[\tilde D^4_B(1324)+\tilde D^4_B(1423)+\tilde D^4_B(1342)+\tilde D^4_B(1243)]\Big]\Bigg]\nonumber\\
&+&\frac{16}{N^4}\Bigg[t^{a_4a_3a_2a_1}\tilde D^4_B(1234)+t^{a_3a_4a_2a_1}\tilde D^4_B(1243)+t^{a_4a_2a_3a_1}\tilde D^4_B(1324)\nonumber\\
&&+t^{a_2a_4a_3a_1}\tilde D^4_B(1342)+t^{a_3a_2a_4a_1}\tilde D^4_B(1423)+t^{a_2a_3a_4a_1}\tilde D^4_B(1432)\Bigg]
\end{eqnarray}
Note, that in our normalization $\tilde D^4_P\sim N^2_c$ while $\tilde D^4_B\sim N_c^3$, thus all the terms in eq.(\ref{decomp}) are of the same order in $N_c$.
It is then convenient to represent the reggeization corrections in the following way:
\begin{eqnarray}\label{d4projP}
&&\tilde D^4_P(1234)=D^{4I}_P(1234)-\frac{g^2\,N^3_c}{3}\left[D^2_{14}\delta_{123}+D^2_{14}\delta_{234}+D^2_{12}\delta_{134}+D^2_{13}\delta_{124}\right]+\nonumber\\
&&\hspace{5cm}+\frac{g^2\,N^3}{4}\Big[D^2_{12}\delta_{14}\delta_{23}+D^2_{12}\delta_{13}\delta_{24}+2D^2_{13}\delta_{12}\delta_{34}\Big]\nonumber\\ && \nonumber \\
&&\tilde D^{4}_B(1234)+\tilde D^{4}_B(2143)=D^{4I}_B(1234)+D^{4I}_B(2143)\\
&&\ \ \ \ \ \ \ \ \ \ \ \ \ \ \ \ \ \ \ \ \ \ \ \ \ \ \ \ \ \ \ -\frac{g^2}{3}\frac{N^4}{8}\left[D^2_{14}\delta_{123}+D^2_{14}\delta_{234}+D^2_{12}\delta_{134}+D^2_{13}\delta_{124}\right]+\nonumber\\
&&\hspace{4.8cm}+\frac{g^2}{2}\frac{N^4}{8}\Big[D^2_{12}\delta_{14}\delta_{23}+D^2_{13}\delta_{12}\delta_{34}\Big]\nonumber\label{d4projB}
\end{eqnarray}
The antisymmetric combination $\tilde D^{4}_B(1234)-\tilde D^{4}_B(2143)$ is odd under charge conjugation. As such it vanishes in the virtual photon state, and will not be of interest to us in the following.

Our goal now is to show that the irreducible pieces $D^{4I}_P$ and $D^{4I}_B$ are simply related to the conjugate reggeon operators $P^\dagger P^\dagger$ and $B^\dagger$ respectively.
  
As discussed in the introduction, in the framework of RFT the expectation value of any operator $\cal O$ in a QCD state with a single dipole is given by the matrix element
\begin{equation}
\langle d(x,y)|{\cal O}\rangle\,\equiv \,\langle {\cal O}\rangle_{d(x,y)}\, \equiv\,\int d\rho\delta[\rho]\frac{1}{N_c}{\rm tr}[R(x)R^\dagger(y)]\,{\cal O}[\rho]
\end{equation}
To make direct contact with ref.\cite{BW},\cite{BE} we should consider the virtual photon initial state, which is the signature and charge conjugation even superposition of two dipoles $d(x,y)$ and $d(y,x)$. Recalling our definition of the Pomeron operator $P(x,y)$, we see that the relevant matrix element for the discussion of \cite{BW, BE} is
\begin{equation}
\langle P(x,y)|{\cal O}\rangle\,\equiv\,\langle {\cal O}\rangle_{P(x,y)}\, \equiv\,\int d\rho\delta[\rho] \,P(x,y)\,{\cal O}[\rho]
\end{equation}
and its rapidity evolution. 
Our discussion in the rest of this section is however much more general. The relations we derive are valid for expectation values in an {\it arbitrary} state $W[d,Q...]$ as long as this state is S and C even. Extention to S and C odd states is straightforward, but will not be presented here. Thus in the rest of this section we consider the general matrix elements 
\begin{equation}
\langle W[d,Q...]|{\cal O}\rangle\,\equiv \,\langle {\cal O}\rangle_{W}
\end{equation}
Physically this means that our results apply to the evolution of the QCD amplitudes of an arbitrary $C$ even initial hadronic state, and not just the virtual photon.
In order not to clutter notations we will be using the same $D$-function notations  as in \cite{BE}, but this time defined as
\begin{equation}\label{DW}
D^n_{a_1...a_n}(x_1...x_n)\equiv(-i\,g)^n\langle \hat\rho^{a_1}(x_1)...\hat\rho^{a_n}(x_n)\rangle_W
\end{equation}
and similarly for $\tilde D$, with an arbitrary $C,\ S$ even $W$. All the matrix elements of the color charge densities and the conjugate Reggeon operators are also understood as taken in the same state $W$.

As a first step, we note that the two point function $D^2_{12}$ is almost identical to the matrix element of the conjugate Pomeron operator $P^\dagger_{12}$. To see this explicitly we represent the LHS of eq.(\ref{2j}) in terms of the symmetrized product of two $\rho$'s using the representation eq.(\ref{js})
\begin{equation}
\frac{1}{2N_c}\Bigg[J_L^a(1)J_L^a(2)+J_R^a(1)J_R^a(2)\Bigg]=\frac{1}{N_c}\rho^a(1)\rho^a(2)+...
\end{equation}
The ellipsis denote the terms which have at least one derivative $\delta/\delta\rho$ to the right of $\rho$ and thus vanish upon taking the matrix element in the Pomeron state.
Comparing this with the definition of $D^2$ we conclude
\begin{equation}\label{d2p}
-ND^2_{12}\,=\,g^2\,\langle P^\dagger_{12}-\delta_{12}\int_zP^\dagger_{1z}\rangle
\end{equation}
The extra term on the RHS of eq.(\ref{d2p}) simply ensures that $D^2$ vanishes when integrated over one of its coordinates. 
\begin{equation}
\int_zD^2_{1z}=0
\end{equation}
This relation has to be satisfied by $D^2$, since the integral of color charge density must vanish when acting on any gauge invariant state.
With the account of this additional term one can straightforwardly check that the evolution equation for $P^\dagger$ eq.(\ref{conevol1}) is equivalent to the evolution equation for $D^2$ eq.(\ref{d2ev}).

Now consider the projection of the four point function onto pairwise singlet states $\tilde D^4_P$. It is straightforward to relate it to the matrix element of two conjugate Pomeron operators $P^\dagger_{12}P^\dagger_{34}$ by considering the matrix element of $J^2_{12}J^2_{34}$.
Using eq.(\ref{js}) we can explicitly write this matrix element in terms of $\rho$ and $\delta/\delta\rho$. Since we are interested in matrix elements of the form eq.(\ref{matrrx}), we should order the factors $\rho$ and $\delta/\delta\rho$ so that all the $\delta/\delta\rho$ factors are to the right of $\rho$. Then any term that contains a nonzero number of factors of $\delta/\delta\rho$, does not contribute to the matrix element. The expression for each individual charge density, eq.(\ref{js}) is already ordered in the correct way. However we still have to order the factors of $\rho$ and $\tau\sim\delta/\delta\rho$ that come from different color rotation operators. Since there are only four factors of $\rho$ in the product, and we are interested in terms of order $\rho^4$ and $\rho^2$ it is enough to expand eq.(\ref{js}) to second order in $\tau$:
\begin{equation}
J_L^a(z)=\rho^b(z)[1-\frac{\tau(z)}{2}+\frac{\tau^2(z)}{12}+...]^{ba}; \ \ \ \ J_R^a(z)=\rho^b(z)[1+\frac{\tau(z)}{2}+\frac{\tau^2(z)}{12}+...]^{ba}
\end{equation}
After some simple algebra we obtain
\begin{eqnarray}
&&\frac{1}{4N^2_c}
\Big[J_L^a(1)J_L^a(2)+J_R^a(1)J_R^a(2)\Big]\Big[J_L^b(3)J_L^b(4)+J_R^b(3)J_R^b(4)\Big]
=\frac{1}{N^2}\rho_1^a\rho_2^a\rho_3^b\rho_4^b\\
&&+\frac{1}{6 N}\rho_1^a\rho_2^a\delta_{234}+\frac{1}{6 N}\rho_1^a\rho_2^a\delta_{134}
+\frac{1}{6 N}\rho_1^a\rho_4^a\delta_{123}+\frac{1}{6 N}\rho_1^a\rho_3^a\delta_{124}
-\frac{1}{4N}\rho^a_1\rho^a_2\delta_{13}\delta_{24}\nonumber\\
&&-\frac{1}{4N}\rho^a_1\rho^a_2\delta_{14}\delta_{23}\Big]\nonumber
\end{eqnarray}
or
\begin{eqnarray}\label{j2rho}
\frac{1}{4N^2_c}&&\langle\Big[J_L^a(1)J_L^a(2)+J_R^a(1)J_R^a(2)\Big]\Big[J_L^b(3)J_L^b(4)+J_R^b(3)J_R^b(4)\Big]\rangle=\frac{1}{N^2}
\langle \rho_1^a\rho_2^a\rho_3^b\rho_4^b\rangle\\
&&-\frac{N}{2g^2}\Big[\frac{1}{3}D^2_{12}\delta_{234}+\frac{1}{3}D^2_{12}\delta_{134}
+\frac{1}{3}D^2_{14}\delta_{123}+\frac{1}{3}D^2_{13}\delta_{124}
-\frac{1}{2}D^2_{12}\delta_{13}\delta_{24}-\frac{1}{2}D^2_{12}\delta_{14}\delta_{23}\Big]\nonumber
\end{eqnarray}
On the other hand, using eq.(\ref{2j}) and omitting the terms involving the Odderon we have
\begin{eqnarray}
&&\frac{1}{4N^2_c}\Big[J_L^a(1)J_L^a(2)+J_R^a(1)J_R^a(2)\Big]\Big[J_L^b(3)J_L^b(4)+J_R^b(3)J_R^b(4)\Big]=
P^\dagger_{12}P^\dagger_{34}-\delta_{12}P^\dagger_{34}\int_xP^\dagger_{1x}\nonumber\\&&-\delta_{34}P^\dagger_{12}\int_xP^\dagger_{3x}+\delta_{12}\delta_{34}\int_{x,y}P^\dagger_{1x}P^\dagger_{3y}
-\frac{1}{2}\Big[P^\dagger_{12}\delta_{13}\delta_{24}+P^\dagger_{12}\delta_{14}\delta_{23}+P^\dagger_{13}\delta_{12}\delta_{34}-\delta_{134}P^\dagger_{12}\nonumber\\
&&\hspace{5cm}-\delta_{234}P^\dagger_{12}-\delta_{123}P^\dagger_{14}-\delta_{124}P^\dagger_{13}+\delta_{12}\delta_{34}\delta_{13}\int_xP^\dagger_{1x}\Big]
\end{eqnarray}
Using eq.(\ref{d2p}) this can be written as
\begin{eqnarray}\label{j2p}
&&\frac{1}{4N^2_c}\langle\Big[J_L^a(1)J_L^a(2)+J_R^a(1)J_R^a(2)\Big]\Big[J_L^b(3)J_L^b(4)+J_R^b(3)J_R^b(4)\Big]\rangle=\\
&&\langle P^\dagger_{12}P^\dagger_{34}-\delta_{12}P^\dagger_{34}\int_xP^\dagger_{1x}-\delta_{34}P^\dagger_{12}\int_xP^\dagger_{3x}+\delta_{12}\delta_{34}\int_{x,y}P^\dagger_{1x}P^\dagger_{3y}\rangle\nonumber\\
&&+\frac{N}{2g^2}\Big[D^2_{12}\delta_{13}\delta_{24}+D^2_{12}\delta_{14}\delta_{23}+D^2_{13}\delta_{12}\delta_{34}-\delta_{134}D^2_{12}-\delta_{124}D^2_{12}-\delta_{123}D^2_{14}-\delta_{234}D^2_{13}\Big]\nonumber
\end{eqnarray}
Combining eqs.(\ref{j2rho},\ref{j2p}) gives
\begin{eqnarray}
&&\hspace{-0.3cm}\frac{1}{N^2}\langle\rho_1^a\rho_2^a\rho_3^b\rho_4^b\rangle=\langle P^\dagger_{12}P^\dagger_{34}-\delta_{12}P^\dagger_{34}\int_xP^\dagger_{1x}-\delta_{34}P^\dagger_{12}\int_xP^\dagger_{3x}+\delta_{12}\delta_{34}\int_{x,y}P^\dagger_{1x}P^\dagger_{3y}\rangle\nonumber\\
&&\hspace{-.5cm}-\frac{N}{3g^2}\Big[\delta_{134}D^2_{12}+\delta_{124}D^2_{12}+\delta_{123}D^2_{14}+\delta_{234}D^2_{13}\Big] +\frac{N}{2g^2}D^2_{13}\delta_{12}\delta_{34}+\frac{N}{4g^2}
\Big[D^2_{12}\delta_{13}\delta_{24}+D^2_{12}\delta_{14}\delta_{23}\Big]\nonumber\\
\end{eqnarray}
Now direct comparison with eq.(\ref{d4projP}) 
\begin{equation}\label{d4p2}
\frac{1}{N^2_c}D^{4I}_P(1234)=g^4\langle P^\dagger_{12}P^\dagger_{34}-\delta_{12}P^\dagger_{34}\int_uP^\dagger_{1u}-\delta_{34}P^\dagger_{12}\int_uP^\dagger_{3u}+\delta_{12}\delta_{34}\int_{u,v}P^\dagger_{1u}P^\dagger_{3v}\rangle
\end{equation}
This relation is a natural counterpart to eq.(\ref{d2p}). The irreducible four point function in \cite{BW, BE} was defined by normal ordering with respect to a single Pomeron state. The matrix element of a product of two Pomeron conjugate operators indeed vanishes in a single Pomeron state by construction, thus the proportionality eq.(\ref{d4p2}).  The  last three terms on the right hand side of eq.(\ref{d4p2}) again take care of vanishing of $D^{4I}$ at zero momentum in full analogy with the last term in eq.(\ref{d2p}).

To complete the discussion of the four point function, we have to repeat the same calculation for  $\tilde D^4_B$. 
The symmetrization procedure yields
\begin{eqnarray}
&&\frac{N^3}{16}[J^4_L(1,2,3,4)+J^4_R(1,2,3,4)]=\rho(1,2,3,4)\\
&&+\frac{N^2}{8}\Big[\frac{1}{4}\delta_{12}\delta_{34}\rho^a_1\rho^a_3
-\frac{1}{4}\delta_{14}\delta_{23}\rho^a_1\rho^a_2
+\frac{1}{3}\delta_{123}\rho^a_1\rho^a_4+\frac{1}{3}\delta_{134}\rho^a_1\rho^a_2-\frac{1}{6}\delta_{124}\rho^a_1\rho^a_3-\frac{1}{6}\delta_{234}\rho^a_1\rho^a_2\Big]\nonumber
\end{eqnarray}
The charge conjugation even combination is
\begin{eqnarray}
&&\frac{N^3}{16}[J^4_L(1,2,3,4)+J^4_L(2,1,4,3)+J^4_R(1,2,3,4)+J^4_R(2,1,4,3)]=\rho(1,2,3,4)+\rho(2,1,3,4)\nonumber\\
&&+\frac{N^2}{16}\Big[\delta_{12}\delta_{34}\rho^a_1\rho^a_3-\delta_{14}\delta_{23}\rho^a_1\rho^a_2
+\frac{1}{3}\Big[\delta_{123}\rho^a_1\rho^a_4+\delta_{234}\rho^a_1\rho^a_2+\delta_{124}\rho^a_1\rho^a_3+\delta_{134}\rho^a_1\rho^a_2\Big]\Big]
\end{eqnarray}
Comparing this with eq.(\ref{j4b}) we find
\begin{eqnarray}
&&\tilde D^{4}_B(1234)+\tilde D^{4}_B(2143)=-g^4\frac{N^3}{2}\langle \Big[B^\dagger_{1234}\\
&&-\frac{1}{2}\delta_{12}\int_u\Big(B^\dagger_{1u34}+B^\dagger_{1u43}-B^\dagger_{14u3}\Big)-\frac{1}{2}\delta_{34}\int_u\Big(B^\dagger_{123u}+B^\dagger_{12u3}-B^\dagger_{132u}\Big)\nonumber\\
&&-\frac{1}{2}\delta_{23}\int_u\Big(B^\dagger_{13u4}+B^\dagger_{1u34}-B^\dagger_{134u}\Big)-\frac{1}{2}\int_u\delta_{14}\Big(B^\dagger_{123u}+B^\dagger_{132u}-B^\dagger_{12u3}\Big)\nonumber\\
&&+\delta_{12}\delta_{34}\int_{uv}B^\dagger_{1u3v}+\delta_{14}\delta_{23}\int_uB^\dagger_{1u2v}\Big]\rangle\nonumber\\
&&-\frac{g^2}{3}\frac{N^4}{8}\Big[\delta_{123}D^2_{14}+\delta_{234}D^2_{12}+\delta_{124}D^2_{13}+\delta_{134}D^2_{12}\Big]+\frac{g^2}{2}\frac{N^4}{8}\Big[\delta_{12}\delta_{34}D^2_{13}+\delta_{14}\delta_{23}D^2_{12}\Big]\nonumber
\end{eqnarray}
This yields the natural identification
\begin{eqnarray}
&& D^{4I}_B(1234)+ D^{4I}_B(2143)=-g^4\frac{N^3}{2}\langle \Big[B^\dagger_{1234}\\
&&-\frac{1}{2}\delta_{12}\int_u\Big(B^\dagger_{1u34}+B^\dagger_{1u43}-B^\dagger_{14u3}\Big)-\frac{1}{2}\delta_{34}\int_u\Big(B^\dagger_{123u}+B^\dagger_{12u3}-B^\dagger_{132u}\Big)\nonumber\\
&&-\frac{1}{2}\delta_{23}\int_u\Big(B^\dagger_{13u4}+B^\dagger_{1u34}-B^\dagger_{134u}\Big)-\frac{1}{2}\int_u\delta_{14}\Big(B^\dagger_{123u}+B^\dagger_{132u}-B^\dagger_{12u3}\Big)\nonumber\\
&&+\delta_{12}\delta_{34}\int_{uv}B^\dagger_{1u3v}+\delta_{14}\delta_{23}\int_uB^\dagger_{1u2v}\Big]\rangle\nonumber
\end{eqnarray}

This conclude our discussion of identification of the "`irreducible"' and "`reggeized"' terms. We note that once the identification is made, it is a simple matter to obtain the Bartels' vertex eq.(\ref{vertex}). The evolution of $D^{4I}_P$ follows directly from eq.(\ref{twopomev}), and reproduces eq.(\ref{vertex}). The evolution of $D^{4I}_B$  follows from eq.(\ref{B-reggeonev}). This evolution equation is homogeneous and does not contain a vertex, again in accordance with eq.(\ref{vertex}). 

Thus our finding here is that the reggeized parts as defined in \cite{BW, BE} precisely account for the single conjugate Pomeron terms present in $D^4$, while the irreducible part $D^{4I}$ contains only $P^\dagger P^\dagger$ and $B^\dagger$ terms.

\subsection{Higher point functions.}
Using this template it is in principle straightforward to extend the reggeization procedure to higher point functions. At the level of $D^6$ one should relate the correlator of six color charge densities to the conjugate Reggeons. The actual calculation would require a fair bit of algebra, but the general structure is clear. Schematically, without indicating the color and coordinate indices the relation will have the form
\begin{equation}\label{proj6}
D^6=\langle aX^\dagger+bB^\dagger P^\dagger+cP^\dagger P^\dagger P^\dagger+ dB^\dagger+eP^\dagger P^\dagger+fP^\dagger \rangle
\end{equation}
Here $X^\dagger$ is the conjugate to the X-Reggeon defined (up to appropriate subtraction of $P$ and $B$) in eq.(\ref{X}). 
The functions $a, \ b$ and $c$ are determined by projection of the six point functions onto the irreducible color structures, like in eq.(\ref{decomp}) for $D^4$. The functions $d,\ e$ and $f$ on the other hand arise due to the symmetrization of the charge density factors, as well as the extra terms appearing in relation between the symmetrized correlators and conjugate Reggeons similar to eqs.(\ref{d4projP},\ref{d4projB}). 

The last three terms in eq.(\ref{proj6}) are the reggeized parts of $D^6$. Our discussion in this section defines a strictly algorithmic way of finding these terms without the need for any guesswork, once the basis of Reggeons is chosen.

The irreducible part of $D^6$ is naturally identified with the first three terms
\begin{equation}
D^{6I}=\langle aX^\dagger+bB^\dagger P^\dagger+cP^\dagger P^\dagger P^\dagger\rangle
\end{equation}
The evolution equation for $D^{6I}$ follows directly from the evolution of $X^\dagger$, $B^\dagger$ and $P^\dagger$. It is easy to understand what kind of vertices it must contain. Clearly the original Bartels' vertex $V_{2\rightarrow 4}$ will appear in the evolution of the triple Pomeron piece of $D^{6I}$.

We expect that the evolution of $X^\dagger$ is homogeneous, and does not contain any vertices in the large $N_c$ limit. This is the case for both $P^\dagger$ and $B^\dagger$, and, we believe this feature persists in the large $N_c$ limit for any conjugate Reggeon, which can be written as a single color trace.

Finally, we expect the evolution equation for $P^\dagger B^\dagger$ to contain the vertex $B^\dagger\rightarrow P^\dagger B^\dagger$. This is suggested by the examination of eq.(\ref{conevol}). Multiplication of the last two terms in eq.(\ref{conevol}) by $B^\dagger$ generates a $B^\dagger$ on the RHS of the evolution. Unless there is a miraculous cancelation between these terms and the terms coming from multiplication of eq.(\ref{B-reggeonev}) by $P^\dagger$, such a vertex will appear. To calculate the precise expression for this vertex one needs to keep the terms of the type $B^\dagger P$ on the RHS of eq.(\ref{B-reggeonev}). This exercise is beyond the scope of the present paper. We stress that we expect such vertex to arise already in the large $N_c$ limit, contrary to the hope expressed in \cite{BW, BE} that the only vertex in this limit would be the triple Pomeron vertex $V_{2\rightarrow 4}$.

\section{Discussion and Conclusions.}
We have shown in this paper how to map the KLWMIJ evolution to Reggeon Field Theory.
In the large $N_c$ limit the Reggeon Field theory contains the Pomeron (dipole) degree of freedom and additional degrees of freedom, which in the weak coupling limit reduce to solutions of the BKP hierarchy. The existence of these extra degrees of freedom does not affect the Balitsky-Kovchegov equation of motion satisfied by the Pomeron (dipole) at large $N_c$.

We have discussed in detail equation of motion satisfied by the $B$-reggeon - the four gluon compound state with vacuum quantum numbers. The $B$-reggeon is an important degree of freedom, which contributes on par with the two Pomeron exchange into variety of observables even in the leading $N_c$ approximation\cite{gluons}. An interesting property of this equation, eq.(\ref{bevolution}) is that even in the weak coupling limit it is not homogeneous, but rather contains a source term proportional to the two Pomeron amplitude. This fact has important consequences.

In particular, it is a very interesting and important question, whether there exist a rapidity regime where the contribution of all the Reggeons except for the Pomeron $P$ is parametrically suppressed. If that were the case, the large $N_c$ limit of KLWMIJ would indeed be a simple Pomeron calculus with a single degree of freedom $P$.

Consider first rapidity evolution of the process of a scattering of two perturbatively dilute objects. At the initial rapidity all Reggeon amplitudes are perturbatively small
\begin{equation}
P\propto \alpha_s^2; \ \ \ \ B\propto \alpha_s^4; \ \ \ etc
\end{equation}
where $etc$ refers for higher Reggeon with vacuum quantum numbers. 
In the linear evolution regime the amplitudes with the vacuum quantum numbers grow. The growth of the Pomeron amplitude is given by the BFKL intercept
\begin{equation}
P_Y\propto \alpha_s^2e^{4\ln 2\bar\alpha_sY}
\end{equation}
The situation is a little more complicated for the B-reggeon, since the equation of motion for it even in the linear regime is not homogeneous, but contains a source term proportional to the square of the Pomeron amplitude. The solution of this equation is the sum of the homogeneous solution and a special solution, which is proportional to the source term. The homogenous solution is also an exponentially growing function of rapidity, but with a different intercept
\begin{equation} \label{a}
B_h\propto \alpha_s^4e^{a8\ln 2\bar\alpha_sY}; \ \ \ etc.
\end{equation}
Schematically therefore, the solution for $B$ in the linear regime is of the form
\begin{equation}\label{bsol}
B\propto \alpha_s^4e^{a8\ln 2\bar\alpha_sY}+cP^2
\end{equation}
The Reggeons with nonvacuum quantum numbers, like $O$ and $C$ have to decrease with rapidity, since they can not have a nonvanishing limit at asymptotically large energies (no constant can be signature, or charge conjugation noninvariant).
The constant $a$ in eq.(\ref{a}) is determined by the solution of the BKP equation. The analytic result for the maximal $a$ is not known at present, however in the large $N_c$ limit $a\approx 1/8$, as found  in \cite{korchemsky},\cite{lipatovvega}. For $a<1$ the homogeneous contribution $B_h$ grows with energy slower than the two Pomeron amplitude. At rapidities of order $Y_P=\frac{1}{4\ln 2 \alpha_s}\ln \frac{1}{\alpha^2_s}$ the Pomeron and two Pomeron amplitudes are parametrically of order unity, but $B_h$ is parametrically suppressed by the factor $\alpha_s^{4(1-a)}$. In this rapidity range one could neglect the $B_h$ amplitude relative to the two Pomeron exchange. The B-reggeon is then simply a given function of the two Pomeron amplitude. One should be able to find this function by solving eq.(\ref{bevolution}) keeping only the $P^2$ source on the LHS and neglecting the homogeneous term. If this were the case for all higher reggeons as well (which is indeed true in the large $N_c$ limit \cite{korchemsky},\cite{lipatovvega}), one would have a consistent Pomeron calculus at rapidities of order $Y_P$ where the Pomeron is the only relevant degree of freedom, while all the other fields are "`constrained fields"' and not additional degrees of freedom. 

It is important to realize that even in this regime one cannot neglect the contributions of $B$ to physical observables, but rather express them in terms of the two Pomeron amplitude.
As we have mentioned earlier, the B-reggeon field as we have defined it, is not orthogonal to the two Pomeron state at the initial rapidity. 
It stays so for all energies including in the saturation regime. In fact according to eq.(\ref{B-reggeon}), the field $B$ does not vanish in the saturation regime similarly to $O$ and $C$ but rather has the nonvanishing limit $B=-1$, similarly to the Pomeron field. This follows directly from the definition eq.(\ref{B-reggeon}), since in the saturation regime any field proportional to a product to $R$ matrices, such as $d$ and $Q$, is expected to vanish.
Thus one might hope that there exist a linear combination of $B$ and $P^2$ which is subleading at all energies and thus can be set to zero. This of course would simplify  life tremendously, since one then would obtain a local theory of the Pomeron alone, with "`renormalized"' vertices. 

Naively, a good candidate for such a field is the $\bar B$ constructed in the Appendix. It is by construction orthogonal to $P^2$ at low energies (up to order $(\delta/\delta\rho)^4$). Also from its definition eq.(\ref{B-reggeon1}) it follows, that the field $\bar B$ vanishes in the saturation regime as well. However, in the low density regime the equation for $\bar B$, just like that for $B$, has a source term linear in $P^2$. Thus even though $\bar B$ is orthogonal to $P^2$ at initial rapidity, the two mix in the evolution. Nominally the magnitude  of $\bar B$ at rapidities  of order $Y_P$ is the same as that of $B$, and it cannot be set to zero. Therefore, barring miraculous cancellations, $\bar B$, just like $B$ cannot be neglected in the full kinematic region. The above discussion suggests that the Green's function of $\bar B$ initially grows with the two Pomeron intercept, reaches a maximum in the pre-saturation regime, where the Pomeron is not parametrically small but has not reached saturation yet, and then drops to zero as the saturation regime is reached. The contribution of $\bar B$ remains important in the pre-saturation regime and cannot be neglected.

We were not able to find a {\it local} combination of $B$ and $P^2$ which satisfies equation similar to eq.(\ref{bevolution}) without a $P^2$ source term.

In the second part of this paper we have addressed the relation of the Reggeon Field Theory to the approach of Bartels et.al. \cite{BW,BE}. We have shown that the prescription of \cite{BE} to split the color charge density correlation functions $D^n$ into irreducible and reggeized parts finds a very natural place in the Reggeon Field Theory framework. The irreducible part corresponds to extraction of the contribution of the highest reggeon (or rather reggeon conjugate momentum) to $D^n$, while the reggeized parts correspond to the contributions of reggeon conjugates of lower dimensionality. This splitting is unambiguously defined, and can be algorithmically implemented for higher correlation functions $D^n$ than those considered in \cite{BW,BE}.
The Bartels' vertex appears in the evolution of  the square of the conjugate Pomeron operator( $P^{\dagger 2}$), and is generated by the triple Pomeron term $P^2P^\dagger$ in the RFT Hamiltonian. The very same term generates the nonlinear interaction term in the Balitsky-Kovchegov equation for the Pomeron. In this sense the Bartels' vertex  does not carry independent information additional to the Balitsky-Kovchegov equation.

Finally we want to make several comments regarding extensions of our approach.

 First, all calculations of the present paper were performed in the large $N_c$ limit. Extension beyond the large $N_c$ in principle is not difficult. It was shown in ref.(\cite{remarks}) how to achieve this in the restricted case when only the Pomeron and Odderon (dipole) degrees of freedom are considered. When one acts with the KLWMIJ Hamiltonian on $W[d]$, the two color charge densities $J_{L(R)}$ can act on either the same dipole, or on two different dipoles. Action on the same dipole leads to the Hamiltonian which is linear in the conjugate momenta $P^\dagger$ and $O^\dagger$ as discussed in the present paper. The term where the two $J$'s act on different dipoles is $1/N_c^2$ suppressed, and is quadratic in the conjugate Reggeon operators. It is explicitly given by
\begin{eqnarray}\label{cc}
\delta H_{P,O}&=&\frac{1}{N_c^2}\frac{\bar \alpha_s}{2\pi}\int_{x,y,u,v,z}L_{x,y,u,v;z}
\left[Q_{xyuv}-\, X_{x,y,z,u,v,z}\right]\,
\frac{\delta^2}{\delta d(u,v)\delta d(p,r)}\\
&=&\frac{1}{N_c^2}\frac{\bar \alpha_s}{2\pi}\int_{u,v,p,r,z}L(u,v,p,r,z)
\Big[Q_{xyuv}-\, X_{x,y,z,u,v,z}\Big]\,
\Big[\left(P^\dagger_{xy}-O^\dagger_{xy}\right)\left(P^\dagger_{uv}-O^\dagger_{uv}\right)\Big]\nonumber
\end{eqnarray}
Interestingly, this additional piece in the Hamiltonian contains a number of $1/N^2_c$ suppressed "`merging"' vertices. To find those explicitly we would need to "`project"' the X-reggeon onto states perturbatively orthogonal to $P$ and $B$, analogously to what we did for $Q$ earlier. Assuming that this procedure goes through, we expect the merging vertices of the form  $BP^\dagger P^\dagger$; $BO^\dagger O^\dagger$, $D^-P^\dagger O^\dagger$ and also vertices involving a single $X$-reggeon and two of the conjugates $P^\dagger$ and $O^\dagger$.  

This is worth noting, in view of the folklore that insists that the KLWMIJ Hamiltonian contains only splitting vertices. This folklore, as we see  is true only in the large $N_c$ limit, whereas $O(1/N_c^2)$ suppressed merging vertices are indeed contained in $H_{KLWMIJ}$.

It is clear that allowing $W$ to depend on all Reggeons will generate additional triple Reggeon merging vertices. These vertices can be obtained in the same way exactly, turning the action of $J_{L(R)}$ on $W$ into functional derivatives with respect to the Reggeon fields, and therefore into conjugate Reggeons. We expect that all triple Reggeon vertices of the form $R_1R^\dagger_2R^\dagger_3$ that are allowed by discrete symmetries will appear.
 The triple Pomeron vertex we have re-derived is known to posses a 2d conformal
invariance \cite{Korchemsky,BLW}. The conformal invariance of the triple Pomeron vertex presumably follows
from the conformal invariance of the JIMWLK/KLWMIJ Hamiltonians. We therefore expect
this symmetry to be present in all Reggeon vertices.

Another interesting direction into which our results should be generalized, is including the Pomeron (Reggeon) loops. The KLWMIJ evolution is valid as long as the projectile is dilute, but the target is dense. The opposite regime of the dilute target and dense projectile is described by the JIMWLK evolution.
Our construction can be immediately transponded into that regime. The conjugate Reggeons are now nonlinear functions of the color charge density, and their explicit form is obtain via the dense-dilute duality transformation \cite{duality} $\delta/\delta\rho^a(x)\rightarrow i\alpha^a(x)$. The natural objects in the JIMWLK regime, are not
 the conjugate Reggeons themselves but the dual Reggeons, for example
\begin{equation}
\bar P_{xy}=\frac{1}{2N_c}\Big[2-{\rm Tr} S(x)S^\dagger (y)-{\rm Tr} S(y)S^\dagger (x)\Big]; \ \ \ \ \ \ \bar O_{xy}=\frac{1}{2N_c}\Big[{\rm Tr} S(x)S^\dagger (y)-{\rm Tr} S(y)S^\dagger (x)\Big];
\end{equation}
where
\begin{equation}
S(x)=\exp\{iT^a\alpha^a(x)\}
\end{equation}
The projectile color field $\alpha^a(x)$ is related to the projectile color charge density via the classical Yang-Mills equations of motion.
In the dilute projectile limit 
\begin{equation}
\alpha^a(x)=\int_y\frac{1}{\nabla^2_{xy}}\rho^a(y)
\end{equation}
Thus in this limit
there is a simple relation between the dual and conjugate Reggeons, i.e.
\begin{equation}
\bar P_{xy}=\frac{\alpha_s}{4\pi}\int_{uv}\frac{1}{\nabla^2_{xu}\nabla^2_{yv}}P^\dagger_{uv};  \ \ \bar O_{xy}=\frac{\alpha_s}{4\pi}\int_{uv}\frac{1}{\nabla^2_{xu}\nabla^2_{yv}}O^\dagger_{uv}
\end{equation}
and similarly for higher Reggeons.

In the JIMWLK regime, while the conjugate Reggeons are nonlinear in the color charge density, the Reggeons themselves are simple homogeneous functions of $\delta/\delta\rho$, obtained by expanding the expressions eqs.(\ref{pomeron},\ref{odderon}) etc. to leading order.

Several years ago a considerable effort was invested to derive the generalization of the KLWMIJ/JIMWLK hamiltonian, which would include the so called Pomeron loop effects, namely would perform the proper nonlinear resummation in both the Reggeon and the conjugate Reggeon variables. A Hamiltonian that partially achieves this goal has been derived in \cite{aklp}. It would be very interesting to extend the approach of the present paper to derive the multireggeon vertices of this extended Hamiltonian.

\section{Appendix}
\subsection{The B-reggeon amended.}
We seek a field that is a linear combination of $Q$, $P$ and $P^2$, so that in the leading order in $\delta/\delta\rho$ it interpolates only the four gluon BKP states and has no overlap with the square of the Pomeron.
 
To find the appropriate operator we consider the $O(\delta/\delta\rho)^4$ term in $Q$, where all the operators $\tau$ are in different points
\begin{equation}\label{q4}
Q(1,2,3,4)|_{(\delta/\delta\rho)^4}=\frac{1}{N_c}{\rm tr}(T^aT^bT^cT^d)\frac{\delta}{\delta\rho^a_1}\frac{\delta}{\delta\rho^b_2}\frac{\delta}{\delta\rho^c_3}\frac{\delta}{\delta\rho^d_4}
\end{equation}
We subtract from the color tensor in eq.(\ref{q4}) a combination of singlet projectors such that the resulting tensor vanishes when contracted over any pair of indices.
\begin{eqnarray}
&&{\rm tr}(T^aT^bT^cT^d)\rightarrow{\rm tr}(T^aT^bT^cT^d)-\frac{1}{4N_c^2}\frac{N_c^6-2N_c^4-2N_c^2+3}{N_c^2(N_c^4-N_c^2-2)}\left[\delta^{ab}\delta^{cd}+\delta^{ad}\delta^{bc}\right]\nonumber\\
&&\hspace{5.5cm}+\frac{3N_c^2-1}{4N_c^2(N_c^4-N_c^2-2)}\delta^{ac}\delta^{bd}\nonumber\\
&&\hspace{2.5cm}\rightarrow_{N_c\rightarrow\infty} {\rm tr}(T^aT^bT^cT^d)-\frac{1}{4N_c^2}\left[\delta^{ab}\delta^{cd}+\delta^{ad}\delta^{bc}\right]
\end{eqnarray}
In a similar way one has to take care of the terms where some of the coordinates of $\tau$ coincide. The subtractions also have to be written in terms of products of Pomerons in order to preserve the invariance under the full $SU_L(N_c)\times SU_R(N_c)$ group. 
After some algebra, the requisite combination is found to be
\begin{eqnarray}\label{B-reggeon1}
\bar B(1,2,3,4)&=&\frac{1}{4}\left[4-Q(1,2,3,4)-Q(4,1,2,3)-Q(3,2,1,4)-Q(2,1,4,3)\right]\\
&-&\left[P_{12}+P_{14}+P_{23}+P_{34}-P_{13}-P_{24}\right]+\left[P_{12}P_{34}+P_{14}P_{23}\right]\nonumber\\
&+&\frac{1}{2}\left[P_{12}P_{23}+P_{23}P_{34}+P_{14}P_{34}+P_{12}P_{14}+P_{13}^2+P_{24}^2\right]\nonumber\\
&-&\frac{1}{2}\left[P_{13}P_{12}+P_{12}P_{24}+P_{13}P_{34}+P_{13}P_{23}+P_{23}P_{24}+P_{13}P_{14}+P_{24}P_{34}+P_{14}P_{24}\right]\nonumber
\end{eqnarray}
\subsection{The calculation of the conjugate B-reggeon.}
To leading order in $1/N_c$ expansion we find:
\begin{equation}\label{4j}
{\rm tr}(T^aT^bT^cT^d)J_L^a(1)J_L^b(2)J^c_L(3)J_L^d(4)B_{1234}=-\frac{N_c^3}{64}\Lambda^8\Big[d_{32}d_{14}+d_{21}d_{43}\Big]
\end{equation}
and
\begin{eqnarray}
{\rm tr}(T^aT^bT^cT^d)J_L^a(1)J_L^b(1)J^c_L(3)J_L^d(4)B_{1234}&=&\frac{N_c^3}{64}\Lambda^8d_{32}d_{14}\\
{\rm tr}(T^aT^bT^cT^d)J_L^a(1)J_L^b(1)J^c_L(3)J_L^d(4)B_{1243}&=&\frac{N_c^3}{64}\Lambda^8d_{12}d_{43}\nonumber\\
{\rm tr}(T^aT^bT^cT^d)J_L^a(1)J_L^b(1)J^c_L(3)J_L^d(4)B_{1324}&=-&\frac{N_c^3}{64}\Lambda^8d_{14}d_{23}\nonumber\\
{\rm tr}(T^aT^bT^cT^d)J_L^a(1)J_L^b(2)J^c_L(3)J_L^d(1)B_{1234}&=&\frac{N_c^3}{64}\Lambda^8d_{21}d_{43}\nonumber\\
{\rm tr}(T^aT^bT^cT^d)J_L^a(1)J_L^b(2)J^c_L(3)J_L^d(1)B_{1243}&=-&\frac{N_c^3}{64}\Lambda^8d_{21}d_{34}\nonumber\\
{\rm tr}(T^aT^bT^cT^d)J_L^a(1)J_L^b(2)J^c_L(3)J_L^d(1)B_{1324}&=&\frac{N_c^3}{64}\Lambda^8d_{32}d_{41}\nonumber\\
{\rm tr}(T^aT^bT^cT^d)J_L^a(1)J_L^b(2)J^c_L(2)J_L^d(4)B_{1234}&=&\frac{N_c^3}{64}\Lambda^8d_{21}d_{43}\nonumber\\
{\rm tr}(T^aT^bT^cT^d)J_L^a(1)J_L^b(2)J^c_L(2)J_L^d(4)B_{1243}&=&-\frac{N_c^3}{64}\Lambda^8d_{21}d_{34}\nonumber\\
{\rm tr}(T^aT^bT^cT^d)J_L^a(1)J_L^b(2)J^c_L(2)J_L^d(4)B_{1324}&=&\frac{N_c^3}{64}\Lambda^8d_{23}d_{14}\nonumber\\
{\rm tr}(T^aT^bT^cT^d)J_L^a(1)J_L^b(2)J^c_L(3)J_L^d(3)B_{1234}&=&\frac{N_c^3}{64}\Lambda^8d_{32}d_{14}\nonumber\\
{\rm tr}(T^aT^bT^cT^d)J_L^a(1)J_L^b(2)J^c_L(3)J_L^d(3)B_{1243}&=&\frac{N_c^3}{64}\Lambda^8d_{21}d_{34}\nonumber\\
{\rm tr}(T^aT^bT^cT^d)J_L^a(1)J_L^b(2)J^c_L(3)J_L^d(3)B_{1324}&=&-\frac{N_c^3}{64}\Lambda^8d_{41}d_{32}\nonumber\\
{\rm tr}(T^aT^bT^cT^d)J_L^a(1)J_L^b(2)J^c_L(1)J_L^d(4)B_{ijkl}&=&{\rm tr}(T^aT^bT^cT^d)J_L^a(1)J_L^b(2)J^c_L(3)J_L^d(2)B_{ijkl}=0\nonumber
\end{eqnarray}
and
\begin{eqnarray}
{\rm tr}(T^aT^bT^cT^d)J_L^a(1)J_L^b(1)J^c_L(3)J_L^d(3)B_{1234}&=-&\frac{N_c^3}{64}\Lambda^8\Big[\Big(d_{12}d_{34}+d_{32}d_{14}\Big)\Big]\nonumber\\
{\rm tr}(T^aT^bT^cT^d)J_L^a(1)J_L^b(1)J^c_L(3)J_L^d(3)B_{1243}&=&0\nonumber\\
{\rm tr}(T^aT^bT^cT^d)J_L^a(1)J_L^b(1)J^c_L(3)J_L^d(3)B_{1324}&=&0\nonumber\\
{\rm tr}(T^aT^bT^cT^d)J_L^a(1)J_L^b(2)J^c_L(2)J_L^d(1)B_{1234}&=-&\frac{N_c^3}{64}\Lambda^8\Big[\Big(Q_{2341}+d_{21}d_{43}-2d_{21}\Big)\Big]\nonumber\\
{\rm tr}(T^aT^bT^cT^d)J_L^a(1)J_L^b(2)J^c_L(2)J_L^d(1)B_{1243}&=-&\frac{N_c^3}{64}\Lambda^8\Big[\Big(Q_{2431}+d_{21}d_{34}-2d_{21}\Big)\Big]\nonumber\\
{\rm tr}(T^aT^bT^cT^d)J_L^a(1)J_L^b(2)J^c_L(2)J_L^d(1)B_{1324}&=-&\frac{N_c^3}{64}\Lambda^82d_{21}\nonumber\\
{\rm tr}(T^aT^bT^cT^d)J_L^a(1)J_L^b(2)J^c_L(1)J_L^d(2)B_{1234}&=&0\nonumber\\
{\rm tr}(T^aT^bT^cT^d)J_L^a(1)J_L^b(2)J^c_L(1)J_L^d(2)B_{1243}&=&0\nonumber\\
{\rm tr}(T^aT^bT^cT^d)J_L^a(1)J_L^b(2)J^c_L(1)J_L^d(2)B_{1324}&=&0\nonumber
\end{eqnarray}
and
\begin{eqnarray}
{\rm tr}(T^aT^bT^cT^d)J_L^a(1)J_L^b(1)J^c_L(1)J_L^d(4)B_{1234}&=&\frac{N_c^3}{64}\Lambda^8\Big[Q_{1234}+d_{14}d_{32}-2d_{14}\Big]\\
{\rm tr}(T^aT^bT^cT^d)J_L^a(1)J_L^b(1)J^c_L(1)J_L^d(4)B_{1324}&=&\frac{N_c^3}{64}\Lambda^8\Big[Q_{1324}+d_{14}d_{23}-2d_{14}\Big]\nonumber\\
{\rm tr}(T^aT^bT^cT^d)J_L^a(1)J_L^b(1)J^c_L(1)J_L^d(4)B_{1243}&=&-\frac{N_c^3}{64}\Lambda^8\Big[d_{12}d_{43}+d_{13}d_{42}-2d_{14}\Big]\nonumber\\
{\rm tr}(T^aT^bT^cT^d)J_L^a(1)J_L^b(1)J^c_L(1)J_L^d(1)B_{1234}&=&-\frac{N_c^3}{64}\Lambda^8\Big[Q_{1234}+Q_{3214}-2\Big(d_{12}+d_{14}-d_{13}\Big)\Big]\nonumber\\
{\rm tr}(T^aT^bT^cT^d)J_L^a(1)J_L^b(1)J^c_L(1)J_L^d(1)B_{1243}&=&-\frac{N_c^3}{64}\Lambda^8\Big[Q_{1243}+Q_{4213}-2\Big(d_{12}+d_{13}-d_{14}\Big)\Big]\nonumber\\
{\rm tr}(T^aT^bT^cT^d)J_L^a(1)J_L^b(1)J^c_L(1)J_L^d(1)B_{1324}&=&-\frac{N_c^3}{64}\Lambda^8\Big[Q_{1324}+Q_{2314}-2\Big(d_{13}+d_{14}-d_{12}\Big)\Big]\nonumber
\end{eqnarray}
and
\begin{equation}
{\rm tr}(T^aT^bT^cT^d)J_L^a(1)J_L^b(2)J^c_L(3)J_L^d(4)B_{1324}={\rm tr}(T^aT^bT^cT^d)J_L^a(1)J_L^b(2)J^c_L(3)J_L^d(4)B_{1342}=0
\end{equation}
And
\begin{equation}
{\rm tr}(T^aT^bT^cT^d)J_L^a(1)J_L^b(2)J^c_L(3)J_L^d(4)C_{1234}=-\frac{N_c^3}{64}\Lambda^8\Big[d_{32}d_{14}+d_{21}d_{43}\Big]
\end{equation}
\begin{eqnarray}
&&{\rm tr}(T^aT^bT^cT^d)J_L^a(1)J_L^b(1)J^c_L(3)J_L^d(4)C_{1234}=\frac{N_c^3}{64}d_{32}d_{14}\\
&&{\rm tr}(T^aT^bT^cT^d)J_L^a(1)J_L^b(1)J^c_L(3)J_L^d(4)C_{1243}=-\frac{N_c^3}{64}\Lambda^8d_{12}d_{43}\nonumber\\
&&{\rm tr}(T^aT^bT^cT^d)J_L^a(1)J_L^b(1)J^c_L(3)J_L^d(4)C_{1324}=-\frac{N_c^3}{64}\Lambda^8d_{14}d_{23}\nonumber\\
&&{\rm tr}(T^aT^bT^cT^d)J_L^a(1)J_L^b(2)J^c_L(3)J_L^d(1)C_{1234}=\frac{N_c^3}{64}\Lambda^8d_{21}d_{43}\nonumber\\
&&{\rm tr}(T^aT^bT^cT^d)J_L^a(1)J_L^b(2)J^c_L(3)J_L^d(1)C_{1243}=-\frac{N_c^3}{64}\Lambda^8d_{21}d_{34}\nonumber\\
&&{\rm tr}(T^aT^bT^cT^d)J_L^a(1)J_L^b(2)J^c_L(3)J_L^d(1)C_{1324}=-\frac{N_c^3}{64}\Lambda^8d_{32}d_{41}\nonumber\\
&&{\rm tr}(T^aT^bT^cT^d)J_L^a(1)J_L^b(2)J^c_L(2)J_L^d(4)C_{1234}=\frac{N_c^3}{64}\Lambda^8d_{21}d_{43}\nonumber\\
&&{\rm tr}(T^aT^bT^cT^d)J_L^a(1)J_L^b(2)J^c_L(2)J_L^d(4)C_{1243}=-\frac{N_c^3}{64}\Lambda^8d_{21}d_{34}\nonumber\\
&&{\rm tr}(T^aT^bT^cT^d)J_L^a(1)J_L^b(2)J^c_L(2)J_L^d(4)C_{1324}=\frac{N_c^3}{64}\Lambda^8d_{23}d_{14}\nonumber\\
&&{\rm tr}(T^aT^bT^cT^d)J_L^a(1)J_L^b(2)J^c_L(3)J_L^d(3)C_{1234}=\frac{N_c^3}{64}\Lambda^8d_{32}d_{14}\nonumber\\
&&{\rm tr}(T^aT^bT^cT^d)J_L^a(1)J_L^b(2)J^c_L(3)J_L^d(3)C_{1243}=\frac{N_c^3}{64}\Lambda^8d_{21}d_{34}\nonumber\\
&&{\rm tr}(T^aT^bT^cT^d)J_L^a(1)J_L^b(2)J^c_L(3)J_L^d(3)C_{1324}=\frac{N_c^3}{64}\Lambda^8d_{41}d_{32}\nonumber\\
&&{\rm tr}(T^aT^bT^cT^d)J_L^a(1)J_L^b(2)J^c_L(1)J_L^d(4)C_{ijkl}={\rm tr}(T^aT^bT^cT^d)J_L^a(1)J_L^b(2)J^c_L(3)J_L^d(2)C_{ijkl}=0\nonumber
\end{eqnarray}
and
\begin{eqnarray}
{\rm tr}(T^aT^bT^cT^d)J_L^a(1)J_L^b(1)J^c_L(3)J_L^d(3)C_{1234}&=&\frac{N_c^3}{64}\Lambda^8\Big[\Big(d_{12}d_{34}-d_{32}d_{14}\Big)\Big]\nonumber\\
{\rm tr}(T^aT^bT^cT^d)J_L^a(1)J_L^b(1)J^c_L(3)J_L^d(3)C_{1243}&=&0\nonumber\\
{\rm tr}(T^aT^bT^cT^d)J_L^a(1)J_L^b(1)J^c_L(3)J_L^d(3)C_{1324}&=&0\nonumber\\
{\rm tr}(T^aT^bT^cT^d)J_L^a(1)J_L^b(2)J^c_L(2)J_L^d(1)C_{1234}&=+&\frac{N_c^3}{64}\Lambda^8\Big[\Big(Q_{2341}-d_{21}d_{43}\Big)\Big]\nonumber\\
{\rm tr}(T^aT^bT^cT^d)J_L^a(1)J_L^b(2)J^c_L(2)J_L^d(1)C_{1243}&=+&\frac{N_c^3}{64}\Lambda^8\Big[\Big(Q_{2431}-d_{21}d_{34}\Big)\Big]\nonumber\\
{\rm tr}(T^aT^bT^cT^d)J_L^a(1)J_L^b(2)J^c_L(2)J_L^d(1)C_{1324}&=&0\nonumber\\
{\rm tr}(T^aT^bT^cT^d)J_L^a(1)J_L^b(2)J^c_L(1)J_L^d(2)C_{1234}&=&0\nonumber\\
{\rm tr}(T^aT^bT^cT^d)J_L^a(1)J_L^b(2)J^c_L(1)J_L^d(2)C_{1243}&=&0\nonumber\\
{\rm tr}(T^aT^bT^cT^d)J_L^a(1)J_L^b(2)J^c_L(1)J_L^d(2)C_{1324}&=&0\nonumber
\end{eqnarray}
and
\begin{eqnarray}
{\rm tr}(T^aT^bT^cT^d)J_L^a(1)J_L^b(1)J^c_L(1)J_L^d(4)C_{1234}&=&-\frac{N_c^3}{64}\Lambda^8\Big[Q_{1234}-d_{14}d_{32}\Big]\\
{\rm tr}(T^aT^bT^cT^d)J_L^a(1)J_L^b(1)J^c_L(1)J_L^d(4)C_{1324}&=&-\frac{N_c^3}{64}\Lambda^8\Big[Q_{1324}-d_{14}d_{23}\Big]\nonumber\\
{\rm tr}(T^aT^bT^cT^d)J_L^a(1)J_L^b(1)J^c_L(1)J_L^d(4)C_{1243}&=&-\frac{N_c^3}{64}\Lambda^8\Big[d_{13}d_{42}-d_{12}d_{43}\Big]\nonumber
\end{eqnarray}
and
\begin{eqnarray}
{\rm tr}(T^aT^bT^cT^d)J_L^a(1)J_L^b(1)J^c_L(1)J_L^d(1)C_{1234}&=&\frac{N_c^3}{64}\Lambda^8\Big[Q_{1234}-Q_{3214}\Big]\nonumber\\
{\rm tr}(T^aT^bT^cT^d)J_L^a(1)J_L^b(1)J^c_L(1)J_L^d(1)B_{1243}&=&\frac{N_c^3}{64}\Lambda^8\Big[Q_{1243}-Q_{4213}\Big]\nonumber\\
{\rm tr}(T^aT^bT^cT^d)J_L^a(1)J_L^b(1)J^c_L(1)J_L^d(1)B_{1324}&=&\frac{N_c^3}{64}\Lambda^8\Big[Q_{1324}-Q_{2314}\Big]\nonumber\\
{\rm tr}(T^aT^bT^cT^d)J_L^a(1)J_L^b(1)J^c_L(3)J_L^d(1)C_{1234}&=&0\nonumber\\
{\rm tr}(T^aT^bT^cT^d)J_L^a(1)J_L^b(2)J^c_L(3)J_L^d(4)C_{1324}&=&{\rm tr}(T^aT^bT^cT^d)J_L^a(1)J_L^b(2)J^c_L(3)J_L^d(4)C_{1342}=0\nonumber\\
\end{eqnarray}
For similar products of right charges we find the same expressions with the raight hand side transformed $R\rightarrow R^\dagger$, which translates into $d_{ij}\rightarrow d_{ji}; 
 \ Q_{ijkl}\rightarrow Q_{jkli}$.

It is easy to check that 
\begin{equation}
{\rm tr}(T^aT^bT^cT^d)J_L^a(1)J_L^b(2)J^c_L(3)J_L^d(4)P_{12}P_{34}=O(N_c)
\end{equation}
and is therefore subleading in $N_c$ relative to the leading terms in eq.(\ref{4j}).

One does get a nonvanishing result when acting on a single Pomeron and an Odderon. This contribution does not involve the conjugate B-reggeon operators, and can be calculated easily to yield
\begin{eqnarray}
&&{\rm tr}(T^aT^bT^cT^d)J_L^a(1)J_L^b(2)J^c_L(3)J_L^d(4)_{(P^\dagger,O^\dagger)}=\frac{N^3}{16}\Big[-\delta_{14}\delta_{23}(1-P_{12}+O_{21})(P^\dagger_{12}-O^\dagger_{21})\nonumber\\
&&+\delta_{123}(1-P_{14}+O_{14})(P^\dagger_{14}-O^\dagger_{14})+\delta_{234}(1-P_{12}+O_{21})(P^\dagger_{12}-O^\dagger_{21})\nonumber\\
&&-\delta_{1234}\int_u(1-P_{1u}+O_{1u})(P^\dagger_{1u}-O^\dagger_{1u})\Big]
\end{eqnarray}
Taking into account the symmetry properties of $B$ and $C$ ($B_{1234}=B_{2143}=B_{3214}; \ \ C_{1234}=-C_{2143}=-C_{3214}$) and defining for convenience
\begin{equation}
J^4_L(1,2,3,4)\equiv \frac{8}{N_c^3}{\rm tr}(T^aT^bT^cT^d)J_L^a(1)J_L^b(2)J^c_L(3)J_L^d(4)
\end{equation}
we obtain the following relations. We present these relations in full here, although we only use them to the lowest order in $P$ in this paper.
\begin{equation}
\Big[J^4_L(1,2,3,4)+J^4_L(2,1,4,3)+J^4_R(1,2,3,4)+J^4_R(2,1,4,3)\Big]=J_B+J_C+J_P
\end{equation}
with
\begin{eqnarray}
&&J_B=
-2B^\dagger_{1234}\Big[d_{14}d_{32}+d_{41}d_{23}+d_{12}d_{34}+d_{21}d_{43}\Big]\\
&&+\delta_{12}\int_u\Big[ B^\dagger_{1u34}\Big(d_{14}d_{3u}+d_{41}d_{u3}+d_{1u}d_{34}+d_{u1}d_{43}\Big)\nonumber\\
&&+B^\dagger_{1u43}\Big(d_{43}d_{1u}+d_{34}d_{u1}+d_{13}d_{4u}+d_{31}d_{u4}\Big)-B^\dagger_{14u3}\Big(d_{14}d_{u3}+d_{41}d_{3u}+d_{13}d_{u4}+d_{31}d_{4u}\Big)\Big]
\nonumber\\
&&+\delta_{34}\int_u\Big[ B^\dagger_{123u}\Big(d_{32}d_{1u}+d_{23}d_{u1}+d_{3u}d_{12}+d_{u3}d_{21}\Big)\nonumber\\
&&+B^\dagger_{12u3}\Big(d_{21}d_{3u}+d_{12}d_{u3}+d_{31}d_{2u}+d_{13}d_{u2}\Big)-B^\dagger_{132u}\Big(d_{32}d_{u1}+d_{23}d_{1u}+d_{31}d_{u2}+d_{13}d_{2u}\Big)\Big]
\nonumber\\
&&+2\delta_{23}\int_u\Big[ B^\dagger_{13u4}\Big(d_{31}d_{4u}+d_{13}d_{u4}\Big)+B^\dagger_{1u34}\Big(d_{14}d_{3u}+d_{41}d_{u3}\Big)-B^\dagger_{134u}\Big(d_{31}d_{u4}+d_{13}d_{4u}\Big)\Big]
\nonumber\\
&&+2\delta_{14}\int_u\Big[ B^\dagger_{123u}\Big(d_{12}d_{3u}+d_{21}d_{u3}\Big)+B^\dagger_{132u}\Big(d_{32}d_{u1}+d_{23}d_{1u}\Big)-B^\dagger_{12u3}\Big(d_{3u}d_{21}+d_{u3}d_{12}\Big)\Big]
\nonumber\\
&&-2\delta_{12}\delta_{34}\int_{uv}B^\dagger_{1u3v}\Big(d_{1u}d_{3v}+d_{3u}d_{1v}+d_{u1}d_{v3}+d_{u3}d_{v1}\Big)\nonumber\\
&&-\delta_{14}\delta_{23}\int_{uv}\Big[B^\dagger_{12uv}\Big(Q_{v12u}+Q_{u21v}+Q_{12uv}+Q_{21vu}+(d_{21}+d_{12})(d_{vu}+d_{uv})-4d_{21}-4d_{12}\Big)\nonumber\\
&& \hspace{12cm}+4B^\dagger_{1u2v}\Big(d_{21}+d_{12}\Big)\Big]\nonumber\\
&&+2\delta_{123}\int_{uv}\Big[B^\dagger_{1uv4}\Big(Q_{1uv4}+Q_{41uv}+d_{41}d_{uv}+d_{14}d_{vu}-2d_{14}-2d_{41}\Big)\nonumber\\
&&\hspace{8cm}-2B^\dagger_{1u4v}\Big(d_{1u}d_{4v}+d_{u1}d_{v4}-d_{14}-d_{41}\Big)\Big]\nonumber\\
&&+2\delta_{124}\int_{uv}\Big[B^\dagger_{1uv3}\Big(Q_{1uv3}+Q_{31uv}+d_{31}d_{uv}+d_{13}d_{vu}-2d_{13}-2d_{31}\Big)\nonumber\\
&&\hspace{8cm}-2B^\dagger_{1u3v}\Big(d_{1u}d_{3v}+d_{u1}d_{v3}-d_{13}-d_{31}\Big)\Big]\nonumber\\
&&+2\delta_{134}\int_{uv}\Big[B^\dagger_{1uv2}\Big(Q_{1uv2}+Q_{21uv}+d_{21}d_{uv}+d_{12}d_{vu}-2d_{12}-2d_{21}\Big)\nonumber\\
&&\hspace{8cm}-2B^\dagger_{1u2v}\Big(d_{1u}d_{2v}+d_{u1}d_{v2}-d_{12}-d_{21}\Big)\Big]\nonumber\\
&&+2\delta_{234}\int_{uv}\Big[B^\dagger_{2uv1}\Big(Q_{2uv1}+Q_{12uv}+d_{12}d_{uv}+d_{21}d_{vu}-2d_{21}-2d_{12}\Big)\nonumber\\
&&\hspace{8cm}-2B^\dagger_{2u1v}\Big(d_{2u}d_{1v}+d_{u2}d_{v1}-d_{21}-d_{12}\Big)\Big]\nonumber\\
&&-2\delta_{1234}\int_{uvz}B^\dagger_{1uvz}\Big[Q_{1uvz}+Q_{uvz1}+Q_{vu1z}+Q_{u1zv}-2\Big(d_{1u}+d_{u1}+d_{1z}+d_{z1}-d_{1v}-d_{v1}\Big)\Big]\nonumber
\end{eqnarray}
\begin{eqnarray}
&&J_P=-2\delta_{14}\delta_{23}(P^\dagger_{12}-P_{12}P^\dagger_{12}-O_{21}O^\dagger_{21})-2\delta_{1234}\int_u(P^\dagger_{1u}-P_{1u}P^\dagger_{1u}-O_{1u}O^\dagger_{1u})\\
&&+\Big[\delta_{123}(P^\dagger_{14}-P_{14}P^\dagger_{14}-O_{14}O^\dagger_{14})+\delta_{234}(P^\dagger_{12}-P_{12}P^\dagger_{12}-O_{21}O^\dagger_{21})\nonumber\\
&&+\delta_{124}(P^\dagger_{13}-P_{13}P^\dagger_{13}-O_{13}O^\dagger_{13})
+\delta_{134}(P^\dagger_{23}-P_{23}P^\dagger_{23}-O_{32}O^\dagger_{32}
\Big]\nonumber
\end{eqnarray}
\begin{eqnarray}
&&J_C=\delta_{12}\int_u\Big[ C^\dagger_{1u34}\Big(d_{14}d_{3u}+d_{41}d_{u3}-d_{1u}d_{34}-d_{u1}d_{43}\Big)\\
&&\hspace{2cm}+C^\dagger_{134u}\Big(d_{43}d_{1u}+d_{34}d_{u1}-d_{13}d_{4u}-d_{31}d_{u4}\Big)\nonumber\\
&&\hspace{2cm}-C^\dagger_{13u4}\Big(d_{14}d_{u3}+d_{41}d_{3u}-d_{13}d_{u4}-d_{31}d_{4u}\Big)\Big]
\nonumber\\
&&+\delta_{34}\int_u\Big[ C^\dagger_{123u}\Big(d_{32}d_{1u}+d_{23}d_{u1}-d_{3u}d_{12}-d_{u3}d_{21}\Big)\nonumber\\
&&\hspace{2cm}+C^\dagger_{12u3}\Big(d_{21}d_{3u}+d_{12}d_{u3}-d_{31}d_{2u}-d_{13}d_{u2}\Big)\nonumber\\
&&\hspace{2cm} +C^\dagger_{132u}\Big(d_{32}d_{u1}+d_{23}d_{1u}-d_{31}d_{u2}-d_{13}d_{2u}\Big)\Big]
\nonumber\\
&&+2\delta_{12}\delta_{34}\int_{uv}C^\dagger_{1u3v}\Big(d_{1u}d_{3v}-d_{3u}d_{1v}+d_{u1}d_{v3}-d_{u3}d_{v1}\Big)\nonumber\\
&&+\delta_{14}\delta_{23}\int_{uv}C^\dagger_{12uv}\Big(Q_{v12u}+Q_{12uv}-Q_{u21v}-Q_{21vu}-(d_{12}-d_{21})(d_{uv}-d_{vu})\Big)\nonumber\\
&&-2\delta_{123}\int_{uv}\Big[C^\dagger_{1uv4}\Big(Q_{1uv4}+Q_{41uv}-d_{41}d_{uv}-d_{14}d_{vu}\Big)\nonumber\\
&&\hspace{2cm}+C^\dagger_{1u4v}\Big(d_{1u}d_{4v}+d_{u1}d_{v4}-d_{1v}d_{4u}-d_{v1}d_{u4}\Big)\Big]\nonumber\\
&&-2\delta_{124}\int_{uv}\Big[C^\dagger_{1uv3}\Big(Q_{1uv3}+Q_{31uv}-d_{31}d_{uv}-d_{13}d_{vu}\Big)\nonumber\\
&&\hspace{2cm}+C^\dagger_{1u3v}\Big(d_{1u}d_{3v}+d_{u1}d_{v3}-d_{1v}d_{3u}-d_{v1}d_{u3}\Big)\Big]\nonumber\\
&&-2\delta_{134}\int_{uv}\Big[C^\dagger_{1uv2}\Big(Q_{1uv2}+Q_{21uv}-d_{21}d_{uv}-d_{12}d_{vu}\Big)\nonumber\\
&&\hspace{2cm}+C^\dagger_{1u2v}\Big(d_{1u}d_{2v}+d_{u1}d_{v2}-d_{1v}d_{2u}-d_{v1}d_{u2}\Big)\Big]\nonumber\\
&&-2\delta_{234}\int_{uv}\Big[C^\dagger_{2uv1}\Big(Q_{2uv1}+Q_{12uv}-d_{12}d_{uv}-d_{21}d_{vu}\Big)\nonumber\\
&&\hspace{2cm}+C^\dagger_{2u1v}\Big(d_{2u}d_{1v}+d_{u2}d_{v1}-d_{2v}d_{1u}-d_{v2}d_{u1}\Big)\Big]\nonumber\\
&&+2\delta_{1234}\int_{uvz}C^\dagger_{1uvz}\Big[Q_{1uvz}+Q_{uvz1}-Q_{vu1z}-Q_{u1zv}\Big]\nonumber
\end{eqnarray}
A similar expression can be straightforwardly derived for the charge conjugation odd combination $J^4_L(1,2,3,4)-J^4_L(2,1,4,3)+J^4_R(1,2,3,4)-J^4_R(2,1,4,3)$. Since we do not need it in the present paper, we are not going to present it explicitly.

To leading order in $P$ these expressions simplify considerably
\begin{eqnarray}
&&-\frac{1}{8}\Big[J^4_L(1,2,3,4)+J^4_L(2,1,4,3)+J^4_R(1,2,3,4)+J^4_R(2,1,4,3)\Big]=B^\dagger_{1234}\\
&&-\frac{1}{2}\delta_{12}\int_u\Big(B^\dagger_{1u34}+B^\dagger_{1u43}-B^\dagger_{14u3}\Big)-\frac{1}{2}\delta_{34}\int_u\Big(B^\dagger_{123u}+B^\dagger_{12u3}-B^\dagger_{132u}\Big)\nonumber\\
&&-\frac{1}{2}\delta_{23}\int_u\Big(B^\dagger_{13u4}+B^\dagger_{1u34}-B^\dagger_{134u}\Big)-\frac{1}{2}\int_u\delta_{14}\Big(B^\dagger_{123u}+B^\dagger_{132u}-B^\dagger_{12u3}\Big)\nonumber\\
&&+\delta_{12}\delta_{34}\int_{uv}B^\dagger_{1u3v}+\delta_{14}\delta_{23}\int_uB^\dagger_{1u2v}\nonumber\\
&&+\frac{1}{4}\delta_{14}\delta_{23}P^\dagger_{12}+\frac{1}{4}\delta_{1234}\int_uP^\dagger_{1u}-\frac{1}{8}\Big[\delta_{123}P^\dagger_{14}+\delta_{234}P^\dagger_{12}+\delta_{124}P^\dagger_{13}
+\delta_{134}P^\dagger_{23}\Big]\nonumber
\end{eqnarray}
This is the expression used in the text.
\subsection{Symmetrization}
Let us consider the relation between nonsymmetrized operators $\rho^{a_1}\rho^{a_2}...\rho^{a_n}$ and fully symmetrized operators $\{\rho^{a_1}\rho^{a_2}...\rho^{a_n} \}$. Using the approach discussed in \cite{dipoles} one can express any product of charge density operators in terms of fully symmetrized products.
As in the text we introduce a simplifying notation
\begin{equation}
\rho^{a_1}_{x_1}...\rho^{a_n}_{x_n}\equiv\frac{1}{n!}\Sigma_{P(1,2...n)}\hat\rho^{a_{P_1}}(x_{P_1})...\hat\rho^{a_{P_n}}(x_{P_n})
\end{equation}

 After some algebra one obtains the following expressions (some already appear in \cite{dipoles}).

For $n=2$
\begin{equation}\label{appd2}
\hat\rho^{a_1}_1\hat\rho^{a_2}_2=\rho^{a_1}_1\rho^{a_2}_2+\frac{1}{2}t^{a_2}_{a_1c}\delta_{12}\rho^{c}_1
\end{equation}
where the subscripts $(1,2...n)$ stands for the coordinate and the superscripts $(a_1,a_2...a_n)$ stands for the color indices of the operator $\rho$. 

For $n=3$
\begin{eqnarray}\label{appd3}
\hat\rho^{a_1}_1\hat\rho^{a_2}_2\hat\rho^{a_3}_3&=&\rho^{a_1}_1\rho^{a_2}_2\rho^{a_3}_3+\frac{1}{2}\bigg[t^{a_2}_{a_1c}\delta_{12}\rho^{c}_1\rho^{a_3}_3+t^{a_3}_{a_1c}\delta_{13}\rho^{c}_1\rho^{a_2}_2+t^{a_3}_{a_2c}\delta_{23}\rho^{a_1}_1\rho^{c}_2\bigg]\nonumber\\
&&+\frac{1}{12}\bigg[\{t^{a_2}t^{a_3}\}_{a_1c}-3(t^{a_3}t^{a_1})_{a_2c}\bigg]\delta_{12}\delta_{13}\rho^c_1
\end{eqnarray}
where $\{t^{a_2}t^{a_3}\}_{a_1c}=(t^{a_2}t^{a_3})_{a_1c}+(t^{a_3}t^{a_2})_{a_1c}$.

For $n=4$
\begin{eqnarray}\label{appd4}
&&\hat\rho^{a_1}_1\hat\rho^{a_2}_2\hat\rho^{a_3}_3\hat\rho^{a_4}_4=\rho^{a_1}_1\rho^{a_2}_2\rho^{a_3}_3\rho^{a_4}_4+\frac{1}{2}
\bigg[
t^{a_2}_{a_1c}\delta_{12}\rho^{c}_1\rho^{a_3}_3\rho^{a_4}_4
+t^{a_3}_{a_1c}\delta_{13}\rho^{c}_1\rho^{a_2}_3\rho^{a_4}_4
+t^{a_4}_{a_1c}\delta_{14}\rho^{c}_1\rho^{a_2}_2\rho^{a_3}_3\nonumber\\
&&\hspace{5cm}+t^{a_3}_{a_2c}\delta_{23}\rho^{a_1}_1\rho^{c}_2\rho^{a_4}_4
+t^{a_4}_{a_2c}\delta_{24}\rho^{a_1}_1\rho^{c}_2\rho^{a_3}_3
+t^{a_4}_{a_3c}\delta_{34}\rho^{a_1}_1\rho^{a_2}_2\rho^{c}_3\bigg]\nonumber\\
&&\hspace{2cm}+\frac{1}{12}\bigg[
g^{a_1a_2a_3c}_{12,13}\rho^c_1\rho^{a_4}_4
+g^{a_1a_2a_4c}_{12,14}\rho^c_1\rho^{a_3}_3
+g^{a_1a_3a_4c}_{13,14}\rho^c_1\rho^{a_2}_2
+g^{a_2a_3a_4c}_{23,24}\rho^{a_1}_1\rho^c_2\nonumber\\
&&\hspace{3.7cm}+3t^{a_3}_{a_2c}t^{a_4}_{a_1d}\delta_{23,14}\rho^d_1\rho^c_2
+3t^{a_4}_{a_2c}t^{a_3}_{a_1d}\delta_{24,13}\rho^d_1\rho^c_2
+3t^{a_4}_{a_3c}t^{a_2}_{a_1d}\delta_{12,34}\rho^d_1\rho^c_3\bigg]\nonumber\\
&&\hspace{2cm}+\frac{1}{4!}\bigg[\big(\{t^{a_1}t^{c}\}t^{a_2}\big)_{a_4a_3}
-\big(t^{a_2}\{t^{a_1}t^{c}\}\big)_{a_4a_3}
+\big(t^{a_2}\{t^{c}t^{a_4}\}\big)_{a_1a_3}
-\big(\{t^{a_3}t^{a_4}\}t^{a_1}\big)_{a_2c}\nonumber\\
&&\hspace{4.7cm}+2(t^{a_4}t^{a_2}t^{a_1})_{a_3c}
-(t^{a_4}t^{a_1}t^{a_2})_{a_3c}\bigg]\delta_{12}\delta_{13}\delta_{14}\rho_1^c\
\end{eqnarray}
where $g$ is defined as 
\begin{equation}
g^{abcd}_{xy,zl}=\bigg[\{t^bt^c\}_{ad}-3(t^ct^a)_{bd}\bigg]\delta_{xy,zl}=-4\Big[2d^{abcd}-d^{abdc}-d^{acbd}\Big]\delta_{xy,zl}
\end{equation}
Also
\begin{equation}
t^{b}_{ae}t^{c}_{de}=2\Big[d^{abcd}-d^{acdb}\Big]
\end{equation}

For $n=5$

\begin{equation}\label{appd5}
\hat\rho^{a_1}_1\hat\rho^{a_2}_2\hat\rho^{a_3}_3\hat\rho^{a_4}_4\hat\rho^{a_5}_5=\rho^{a_1}_1\rho^{a_2}_2\rho^{a_3}_3\rho^{a_4}_4\rho^{a_5}_5+\rho^5_4+\rho^5_3+\rho^5_2
\end{equation}
with
\begin{eqnarray}
&&\rho^5_4=\frac{1}{2}
\bigg[
t^{a_2}_{a_1c}\delta_{12}\rho^{c}_1\rho^{a_2}_2\rho^{a_3}_3\rho^{a_4}_4
+t^{a_3}_{a_1c}\delta_{13}\rho^{c}_1\rho^{a_2}_2\rho^{a_4}_4\rho^{a_5}_5\\
&&
+t^{a_4}_{a_1c}\delta_{14}\rho^{c}_1\rho^{a_2}_2\rho^{a_3}_3\rho^{a_5}_5
+t^{a_5}_{a_1c}\delta_{15}\rho^{c}_1\rho^{a_2}_2\rho^{a_3}_3\rho^{a_4}_4
+t^{a_3}_{a_2c}\delta_{23}\rho^{a_1}_1\rho^{c}_2\rho^{a_4}_4\rho^{a_5}_5
+t^{a_4}_{a_2c}\delta_{24}\rho^{a_1}_1\rho^{c}_2\rho^{a_3}_3\rho^{a_5}_5\nonumber\\
&&
+t^{a_5}_{a_2c}\delta_{25}\rho^{a_1}_1\rho^{c}_2\rho^{a_3}_3\rho^{a_4}_4
+t^{a_4}_{a_3c}\delta_{34}\rho^{a_1}_1\rho^{a_2}_2\rho^{c}_3\rho^{a_5}_5
+t^{a_5}_{a_3c}\delta_{35}\rho^{a_1}_1\rho^{a_2}_2\rho^{c}_3\rho^{a_4}_4
+t^{a_5}_{a_4c}\delta_{45}\rho^{a_1}_1\rho^{a_2}_2\rho^{a_3}_3\rho^{c}_4
\bigg]\nonumber
\end{eqnarray}
\begin{eqnarray}
&&\rho^5_3=\frac{1}{12}\bigg[
g^{a_1a_2a_3c}_{12,13}\rho^c_1\rho^{a_4}_4\rho^{a_5}_5
+g^{a_1a_2a_4c}_{12,14}\rho^c_1\rho^{a_3}_3\rho^{a_5}_5
+g^{a_1a_2a_5c}_{12,15}\rho^c_1\rho^{a_3}_3\rho^{a_4}_4
+g^{a_1a_3a_4c}_{13,14}\rho^c_1\rho^{a_2}_2\rho^{a_5}_5\\
&&
+g^{a_1a_3a_5c}_{13,15}\rho^c_1\rho^{a_2}_2\rho^{a_4}_4
+g^{a_1a_4a_5c}_{14,15}\rho^c_1\rho^{a_2}_2\rho^{a_3}_3
+g^{a_2a_3a_4c}_{23,24}\rho^{a_1}_1\rho^{c}_2\rho^{a_5}_5
+g^{a_2a_3a_5c}_{23,25}\rho^{a_1}_1\rho^{c}_2\rho^{a_4}_4\nonumber\\
&&
+g^{a_2a_4a_5c}_{24,25}\rho^{a_1}_1\rho^{c}_2\rho^{a_3}_3
+g^{a_3a_4a_5c}_{34,35}\rho^{a_1}_1\rho^{a_2}_2\rho^{c}_3\nonumber\\
&&
+\frac{1}{4}\bigg[t^{a_3}_{a_2c}\delta_{23}\bigg(t^{a_4}_{a_1d}\delta_{14}\rho^{d}_1\rho^{c}_2\rho^{a_5}_5+t^{a_5}_{a_1d}\delta_{15}\rho^{d}_1\rho^{c}_2\rho^{a_4}_4\bigg)
+t^{a_4}_{a_2c}\delta_{24}\bigg(t^{a_3}_{a_1d}\delta_{13}\rho^{d}_1\rho^{c}_2\rho^{a_5}_5+t^{a_5}_{a_1d}\delta_{15}\rho^{d}_1\rho^{c}_2\rho^{a_3}_3\bigg)\nonumber\\
&&
+t^{a_5}_{a_2c}\delta_{25}\bigg(t^{a_3}_{a_1d}\delta_{13}\rho^{d}_1\rho^{c}_2\rho^{a_4}_4+t^{a_4}_{a_1d}\delta_{14}\rho^{d}_1\rho^{c}_2\rho^{a_3}_3\bigg)
+t^{a_4}_{a_3c}\delta_{34}\bigg(t^{a_2}_{a_1d}\delta_{12}\rho^{d}_1\rho^{c}_3\rho^{a_5}_5+t^{a_5}_{a_1d}\delta_{15}\rho^{d}_1\rho^{a_2}_2\rho^{c}_3\bigg)\nonumber\\
&&
+t^{a_5}_{a_4c}\delta_{45}\bigg(t^{a_2}_{a_1d}\delta_{12}\rho^{d}_1\rho^{c}_3\rho^{a_4}_4+t^{a_4}_{a_1d}\delta_{14}\rho^{d}_1\rho^{a_2}_2\rho^{a_3}_3\bigg)
+t^{a_5}_{a_4c}\delta_{45}\bigg(t^{a_2}_{a_1d}\delta_{12}\rho^{d}_1\rho^{a_3}_3\rho^{c}_4+t^{a_3}_{a_1d}\delta_{13}\rho^{d}_1\rho^{a_2}_2\rho^{c}_4\bigg)\nonumber\\
&&
+\bigg(t^{a_4}_{a_3c}t^{a_5}_{a_2d}\delta_{25,34}+t^{a_5}_{a_3c}t^{a_4}_{a_2d}\delta_{24,35}\bigg)\rho^{a_1}_1\rho^{d}_2\rho^{c}_3+t^{a_5}_{a_4c}t^{a_3}_{a_2d}\delta_{23,45}\rho^{a_1}_1\rho^{d}_2\rho^{c}_4\bigg]\nonumber
\end{eqnarray}
\begin{eqnarray}
&&\rho^5_2=\frac{1}{24}\bigg[
\bigg(t^{a_4}_{a_3c}\{t^ct^{a_5}\}_{a_1d}+t^{a_5}_{a_3c}\{t^ct^{a_4}\}_{a_1d}+t^{a_5}_{a_4c}\{t^ct^{a_3}\}_{a_1d}\nonumber\\
&&\hspace{2cm}+\big(\{t^{a_4}t^{a_5}\}_{a_3c}-3(t^{a_5}t^{a_3})_{a_4c}\big)t^{c}_{a_1d}\bigg)\delta_{13,34,35}\rho^d_1\rho^{a_2}_2\nonumber\\
&&\hspace{2cm}+(2\leftrightarrow 3)+(2\leftrightarrow 3, 3\leftrightarrow 4)+(2\leftrightarrow 3, 3\leftrightarrow 4, 4\leftrightarrow 5)\bigg]\nonumber\\
&&+\frac{1}{24}\bigg[t^{a_3}_{a_2c}\{t^{a_4}t^{a_5}\}_{a_1d}\delta_{14,15,23}+t^{a_4}_{a_2c}\{t^{a_3}t^{a_5}\}_{a_1d}\delta_{13,15,24}+t^{a_5}_{a_2c}\{t^{a_3}t^{a_4}\}_{a_1d}\delta_{13,14,25}\nonumber\\
&&
+t^{a_5}_{a_1d}\big[\{t^{a_3}t^{a_4}\}_{a_2c}-3(t^{a_4}t^{a_2})_{a_3c}\big]\delta_{15,23,24}+T^{a_4}_{a_1d}\big[\{t^{a_3}t^{a_5}\}_{a_2c}-3(t^{a_5}t^{a_2})_{a_3c}\big]\delta_{14,23,25}\nonumber\\
&&+t^{a_3}_{a_1d}\big[\{t^{a_4}t^{a_5}\}_{a_2c}-3(t^{a_5}t^{a_2})_{a_4c}\big]\delta_{13,24,25}\bigg]\rho^{d}_1\rho^{c}_2\nonumber\\
&&
+\frac{1}{24}\bigg[t^{a_4}_{a_3c}\{t^{a_2}t^{a_5}\}_{a_1d}\delta_{12,15,34}+t^{a_5}_{a_3c}\{t^{a_2}t^{a_4}\}_{a_1d}\delta_{12,14,35}+t^{a_2}_{a_1d}\big[\{t^{a_4}t^{a_5}\big\}_{a_3c}\nonumber\\
&&\hspace{2cm}-3(t^{a_5}t^{a_3})_{a_4c}\big]\delta_{12,34,35}\bigg]\rho^d_1\rho^c_3\nonumber\\
&&
+\frac{1}{24}t^{a_5}_{a_4c}\{t^{a_2}t^{a_3}\}\delta_{12,13,45}]\rho^d_1\rho^c_4\nonumber\\
&&+\frac{1}{8}\bigg[t^{a_4}_{a_3c}t^{a_5}_{a_2d}\delta_{25,34}+t^{a_5}_{a_3c}t^{a_4}_{a_2d}\delta_{24,35}\bigg]\bigg[t^{d}_{a_1e}\delta_{12}\rho^e_1\rho^c_3+t^{c}_{a_1e}\delta_{13}\rho^e_1\rho^d_2\bigg]\nonumber\\
&&+\frac{1}{8}t^{a_5}_{a_4c}t^{a_3}_{a_2d}\delta_{23,45}\big[t^d_{a_1e}\delta_{12}\rho^e_1\rho^c_4+t^c_{a_1e}\delta_{14}\rho^e_1\rho^d_2\big]\nonumber\\
&&+\frac{1}{24}\bigg[(t^{a_5}t^{c}t^{a_4})_{a_2a_3}-(t^{a_4}t^{a_1}t^{a_5})_{a_3c}+(t^{a_4}t^{c}t^{a_5})_{a_2a_3}\nonumber\\
&&-(t^{a_5}t^{a_2}t^{a_4})_{a_3c}+(t^{a_3}t^{c}t^{a_5})_{a_2a_4}-(t^{a_5}t^{a_2}t^{a_3})_{a_4c}-[\{t^{a_4}t^{a_5}\}t^{a_2}]_{a_3c}\nonumber\\
&&\hspace{2cm}+3(t^{a_5}t^{a_3}t^{a_2})_{a_4c}\bigg]\delta_{23,24,25}\rho^{a_1}_{1}\rho^c_{2}
\end{eqnarray}

\section*{Acknowledgments}
T.A., C.C. and M.L. thank the Physics Department of the University of Connecticut for hospitality during the visits while the work 
on this project was in progress.  A.K. and T.A. thank  the  Physics Departments of the Ben-Gurion University of the Negev; A.K. also thanks 
  Universidad T\'ecnica Federico Santa Mar\'\i a.
The research  was supported by the DOE grant DE-FG02-92ER40716;  the  Marie Curie Grant  PIRG-GA-2009-256313; the  ISRAELI SCIENCE FOUNDATION grant \#87277111;  the IRSES network "High-Energy QCD for Heavy Ions";  the  Fondecyt (Chile) grants  1100648, 1130599, DGIP 11.11.05; European Research Council grant HotLHC ERC-2001-
StG-279579; Ministerio de Ciencia e Innovac\'\i on of Spain grants FPA2009-06867-E and Consolider-Ingenio 2010 CPAN CSD2007-00042 and by FEDER.

\end{document}